\documentclass{tlp}

\usepackage{fixtitle}

\usepackage{amsmath}
\usepackage{amssymb}
\usepackage{latexsym}
\usepackage{dcolumn}
\usepackage{hhline}
\usepackage{url}

\newcommand*{\cA}{\ensuremath{\mathcal{A}}}

\newtheorem{theorem}{Theorem}
\newtheorem{definition}[theorem]{Definition}
\newtheorem{example}[theorem]{Example}

\newtheorem{proposition}[theorem]{Proposition}
\newtheorem{corollary}[theorem]{Corollary}

\newdimen\pfskip \pfskip=10pt 
\newdimen\smallpfskip \smallpfskip=2pt 
  {\par\vskip\pfskip plus0.5\pfskip minus0.1\pfskip\noindent
   {\bfseries\proofname~{#1}}\enspace\ignorespaces}%
  {\par\addvspace{\pfskip plus0.5\pfskip minus0.1\pfskip}}
\def\proofname{Proof}

\def\qed{\relax\ifmmode\hskip2em \Box\else\unskip\nobreak\hskip1em $\Box$\fi}

\newcommand{\summary}[1]{\textrm{\textbf{\textup{#1}}}} 

\newenvironment{program}
               {
                 \ttfamily
                 \begin{tabular}{l}
               }
               {
                 \end{tabular}
               }

\renewcommand{\emptyset}{\mathord{\varnothing}}

\newcommand*{\wpf}{\mathop{\wp_\mathrm{f}}\nolimits}

\newcommand*{\sseq}{\subseteq}
\newcommand*{\sseqf}{\mathrel{\subseteq_\mathrm{f}}}
\newcommand*{\sslt}{\subset}
\newcommand*{\Sseq}{\supseteq}

\newcommand{\Nsseq}{\nsubseteq}

\newcommand*{\card}{\mathop{\#\!}\nolimits}
\newcommand*{\union}{\cup}
\newcommand*{\bigunion}{\bigcup}
\newcommand*{\inters}{\cap}

\newcommand*{\setdiff}{\setminus}

\newcommand{\sset}[2]{{\renewcommand{\arraystretch}{1.2}
                      \left\{\,#1 \,\left|\,
                               \begin{array}{@{}l@{}}#2\end{array}
                      \right.   \,\right\}}}

\newcommand*{\concat}{\mathbin{.}}

\newcommand*{\fund}[3]{\mathord{#1}\colon#2\rightarrow#3}

\newcommand*{\compose}{\mathbin{\circ}}

\newcommand*{\entails}{\mathrel{\vdash}}

\newcommand*{\piff}{\mathrel{\leftrightarrow}}
\newcommand*{\pimplies}{\mathrel{\rightarrow}}

\newcommand{\st}{\mathrel{.}}
\newcommand{\itc}{\mathrel{:}}
\newcommand{\defrel}[1]{\mathrel{\buildrel \mathrm{def} \over {#1}}}
\newcommand{\defeq}{\defrel{=}}
\newcommand{\defiff}{\defrel{\Longleftrightarrow}}

\newcommand*{\Vars}{\mathord{\mathit{Vars}}}
\newcommand*{\Sig}{\mathord{\mathit{Sig}}}
\newcommand*{\Terms}{\mathord{\mathit{Terms}}}
\newcommand*{\GTerms}{\mathord{\mathit{GTerms}}}
\newcommand*{\LTerms}{\mathord{\mathit{LTerms}}}
\newcommand*{\HTerms}{\mathord{\mathit{HTerms}}}

\newcommand*{\vars}{\mathop{\mathrm{vars}}\nolimits}
\newcommand*{\mvars}{\mathop{\mathrm{mvars}}\nolimits}
\newcommand*{\occlin}{\mathop{\mathrm{occ\_lin}}\nolimits}

\newcommand*{\Bind}{\mathord{\mathit{Bind}}}
\newcommand*{\bs}{\mathord{\mathit{bs}}}

\newcommand*{\RSubst}{\mathord{\mathit{RSubst}}}
\newcommand*{\VSubst}{\mathord{\mathit{VSubst}}}
\newcommand*{\ISubst}{\mathord{\mathit{ISubst}}}

\newcommand*{\dom}{\mathop{\mathrm{dom}}\nolimits}

\newcommand*{\range}{\mathop{\mathrm{range}}\nolimits}

\newcommand*{\rt}{\mathop{\mathrm{rt}}\nolimits}

\newcommand*{\Eqs}{\mathord{\mathit{Eqs}}}
\newcommand*{\mgs}{\mathop{\mathrm{mgs}}\nolimits}

\newcommand*{\HT}{\ensuremath{\mathcal{FT}}}
\newcommand*{\RT}{\ensuremath{\mathcal{RT}}}

\newcommand{\glb}{\mathop{\mathrm{glb}}\nolimits}

\newcommand{\uco}{\mathop{\mathrm{uco}}\nolimits}
\newcommand{\ucoleq}{\mathrel{\sqsubseteq}}

\newcommand{\ucomeet}{\mathbin{\sqcap}}

\newcommand*{\project}{\mathop{\exists\kern-.4em\exists}\nolimits}

\newcommand*{\VI}{\mathord{\mathit{VI}}}
\newcommand*{\SG}{\mathord{\mathit{SG}}}
\newcommand*{\SH}{\mathord{\mathit{SH}}}
\newcommand*{\sh}{\mathord{\mathit{sh}}}

\newcommand*{\sg}{\mathop{\mathrm{sg}}\nolimits}
\newcommand*{\occ}{\mathop{\mathrm{occ}}\nolimits}
\newcommand*{\ssets}{\mathop{\mathrm{ssets}}\nolimits}

\newcommand*{\rel}{\mathop{\mathrm{rel}}\nolimits}
\newcommand*{\irel}{\mathop{\overline{\mathrm{rel}}}\nolimits}
\newcommand*{\bin}{\mathop{\mathrm{bin}}\nolimits}
\newcommand*{\alub}{\mathop{\mathrm{alub}}\nolimits}

\newcommand*{\aproj}{\mathop{\mathrm{aexists}}\nolimits}

\newcommand*{\mgu}{\mathop{\mathrm{mgu}}\nolimits}
\newcommand*{\amgu}{\mathop{\mathrm{amgu}}\nolimits}

\newcommand*{\aunify}{\mathop{\mathrm{aunify}}\nolimits}

\newcommand*{\PSD}{\mathit{PSD}}

\newcommand*{\PS}{\mathord{\mathit{PS}}}

\newcommand*{\SFL}{\mathit{SFL}}
\newcommand*{\sfl}{\mathit{d}}
\newcommand*{\leqSFL}{\mathrel{\leq_{\scriptscriptstyle S}}}
\newcommand*{\alubSFL}{\alub_{\scriptscriptstyle S}}
\newcommand*{\aprojSFL}{\aproj_{\scriptscriptstyle S}}
\newcommand*{\abstrSFL}{\alpha_{\scriptscriptstyle S}}
\newcommand*{\concrSFL}{\gamma_{\scriptscriptstyle S}}
\newcommand*{\botSFL}{\bot_{\scriptscriptstyle S}}
\newcommand*{\amguSFL}{\amgu_{\scriptscriptstyle S}}
\newcommand*{\cyclicreduce}{\mathop{\mathrm{cyclic}}\nolimits}
\newcommand*{\aunifySFL}{\aunify_{\scriptscriptstyle S}}
\newcommand*{\PSDFL}{\SFL_2}

\newcommand*{\fvars}{\mathop{\mathrm{fvars}}\nolimits}
\newcommand*{\gvars}{\mathop{\mathrm{gvars}}\nolimits}

\newcommand*{\lvars}{\mathop{\mathrm{lvars}}\nolimits}

\newcommand*{\ground}{\mathop{\mathrm{ground}}\nolimits}
\newcommand*{\ind}{\mathop{\mathrm{ind}}\nolimits}
\newcommand*{\free}{\mathop{\mathrm{free}}\nolimits}
\newcommand*{\lin}{\mathop{\mathrm{lin}}\nolimits}

\newcommand*{\sharewith}{\mathop{\mathrm{share\_with}}\nolimits}

\newcommand*{\rhoCon}{\mathop{\rho_{\scriptscriptstyle \Con}}}
\newcommand*{\rhoPS}{\mathop{\rho_{\scriptscriptstyle \PS}}}

\newcommand*{\rhoPSD}{\mathop{\rho_{\scriptscriptstyle \PSD}}}

\newcommand*{\rhoF}{\mathop{\rho_{\scriptscriptstyle F}}}
\newcommand*{\rhoL}{\mathop{\rho_{\scriptscriptstyle L}}}

\newcommand*{\Bool}{\mathit{Bool}}

\newcommand*{\Pos}{\mathit{Pos}}

\newcommand*{\Con}{\mathit{Con}}

\newcommand*{\ASub}{\mathsf{ASub}}
\newcommand*{\botASub}{\mathord{\bot_{\mathsf{\scriptscriptstyle ASub}}}}
\newcommand*{\leqASub}{\mathrel{\preceq_{\mathsf{\scriptscriptstyle ASub}}}}
\newcommand*{\binASub}{\bin}
\newcommand*{\soln}{\mathop{\mathrm{soln}}\nolimits}
\newcommand*{\mayshare}[1]{\mathrel{\buildrel {#1} \over {\iff}}}
\newcommand*{\mayshareoroccurtwice}[1]{\mathrel{\buildrel {#1} \over {\longleftrightarrow}}}
\newcommand*{\amguASub}{\amgu_{\scriptscriptstyle \ASub}}
\newcommand*{\abstrASub}{\alpha_{\scriptscriptstyle \ASub}}

\newcommand*{\Pattern}{\mathop{\mathrm{Pattern}}\nolimits}

\newcommand{\china}{\textmd{\textsc{China}}}

\newcommand*{\etal}{et al.}

\usepackage{stmaryrd}

\allowdisplaybreaks

\begin{document}
\title[Correct and Efficient Integration
       of Set-Sharing, Freeness and Linearity]
       {A Correct, Precise and Efficient Integration
       of Set-Sharing, Freeness and Linearity
       for the Analysis of Finite
       and Rational Tree Languages\thanks{The present work has been
  funded by MURST projects ``Automatic Program Certification
    by Abstract Interpretation'',
  ``Abstract Interpretation, type systems and control-flow analysis'', and
  ``Automatic Aggregate- and Number-Reasoning for Computing:
    from Decision Algorithms to Constraint Programming
    with Multisets, Sets, and Maps'';
  by the Integrated Action Italy-Spain
  ``Advanced Development Environments for Logic Programs'';
  by the University of Parma's FIL scientific research project
  (ex 60\%) ``Pure and applied mathematics'';
  and
  by the UK's Engineering and Physical Sciences Research Council (EPSRC)
  under grant M05645.
       }
}
\author[P. M. Hill, E. Zaffanella, and R. Bagnara]{
       {PATRICIA M. HILL}
       \affiliation
       School of Computing,
       University of Leeds,
       Leeds, U.K. \\
       \email{hill@comp.leeds.ac.uk}
       \and
       {ENEA ZAFFANELLA, ROBERTO BAGNARA}
       \affiliation
       Department of Mathematics,
       University of Parma,
       Italy \\
       \email{ \{zaffanella,bagnara\}@cs.unipr.it}
}
\maketitle

\begin{abstract}
It is well-known that freeness and linearity information
positively interact with aliasing information,
allowing both the precision
and the efficiency of the sharing analysis of logic programs
to be improved.
In this paper we present a novel combination of set-sharing
with freeness and linearity information,
which is characterized by an improved abstract unification operator.
We provide a new abstraction function
and prove the correctness of the analysis
for both the finite tree and the rational tree cases.
Moreover,
we show that the same notion of redundant information
as identified in~\cite{BagnaraHZ02TCS,ZaffanellaHB02TPLP}
also applies to this abstract domain combination:
this allows for the implementation of an abstract unification
operator running in polynomial time and achieving the same
precision on all the considered observable properties.
\end{abstract}


\section{Introduction}

Even though the set-sharing domain is, in a sense,
remarkably precise,
more precision is attainable by combining it with other domains.
In particular, freeness and linearity information has received
much attention by the literature on sharing analysis
(recall that a variable is said to be free if it is not bound to
a non-variable term; it is linear if it is not bound to a term
containing multiple occurrences of another variable).

As argued informally by S{\o}ndergaard~\cite{Sondergaard86},
the mutual interaction between linearity and aliasing information
can improve the accuracy of a sharing analysis.
This observation has been formally applied in~\cite{CodishDY91}
to the specification of the abstract $\mgu$ operator
for the domain $\ASub$.
In his PhD thesis~\cite{Langen90th},
Langen proposed a similar integration with linearity,
but for the set-sharing domain.
He has also shown how the aliasing information allows
to compute freeness with a good degree of accuracy
(however, freeness information was not exploited to improve aliasing).
King~\cite{King94} has also shown how a more refined tracking of linearity
allows for further precision improvements.

The synergy attainable from a bi-directional interaction between
aliasing and freeness information was initially pointed out
by Muthukumar and Hermenegildo \cite{MuthukumarH91,MuthukumarH92}.
Since then, several authors considered the integration of set-sharing
with freeness,
sometimes also including additional explicit structural information
\cite{CodishDFB93,CodishDFB96,File94,KingS94}.

Building on the results obtained
in~\cite{Sondergaard86},~\cite{CodishDY91} and~\cite{MuthukumarH91},
but independently from~\cite{Langen90th},
Hans and Winkler~\cite{HansW92} proposed a combined integration
of freeness and linearity information with set-sharing.
Similar combinations have been proposed
in~\cite{BruynoogheC93,BruynoogheCM94,BruynoogheCM94TR}.
From a more pragmatic point of view,
Codish \etal{}~\cite{CodishMBdlBH93,CodishMBdlBH95}
integrate the information captured by the domains of~\cite{Sondergaard86}
and~\cite{MuthukumarH91} by performing the analysis with both
domains at the same time,
exchanging information between the two components at each step.

Most of the above proposals differ in the carrier
of the underlying abstract domain.
Even when considering the simplest domain combinations
where explicit structural information is ignored,
there is no general consensus on the specification
of the abstract unification procedure.
From a theoretical point of view,
once the abstract domain has been related to the concrete one
by means of a Galois connection,
it is always possible to specify the best correct approximation
of each operator of the concrete semantics.
However,
empirical observations suggest that sub-optimal operators
are likely to result in better complexity/precision trade-offs
\cite{BagnaraZH00}.
As a consequence, it is almost impossible to identify
``the right combination'' of variable aliasing
with freeness and linearity information,
at least when practical issues,
such as the complexity of the abstract unification procedure,
are taken into account.

Given this state of affairs,
we will now consider a domain combination
whose carrier is essentially the same as specified
by Langen~\cite{Langen90th} and Hans and Winkler~\cite{HansW92}.
(The same domain combination was also considered
by Bruynooghe \etal{}~\cite{BruynoogheCM94,BruynoogheCM94TR},
but with the addition of compoundness and explicit structural information.)
The novelty of our proposal lies in the specification
of an improved abstract unification procedure,
better exploiting the interaction between sharing and linearity.
As a matter of fact, we provide an example showing that all previous
approaches to the combination of set-sharing with freeness and linearity
are not uniformly more precise than the analysis based
on the $\ASub$ domain~\cite{CodishDY91,King00,Sondergaard86},
whereas such a property is enjoyed by our proposal.

By extending the results of~\cite{HillBZ02TPLP} to this combination,
we provide a new abstraction function
that can be applied to any logic language computing on domains
of syntactic structures, with or without the occurs-check;
by using this abstraction function,
we also prove the correctness of the new abstract unification procedure.
Moreover, we show that the same notion of redundant information
as identified in~\cite{BagnaraHZ02TCS,ZaffanellaHB02TPLP}
also applies to this abstract domain combination.
As a consequence, it is possible to implement an algorithm
for abstract unification running in polynomial time
and still obtain the same precision on all the considered observables:
groundness, independence, freeness and linearity.

This paper is based on~\cite[Chapter~6]{Zaffanella01th},
the PhD thesis of the second author.
In Section~\ref{sec:prelims}, we define some notation
and recall the basic concepts used later in the paper.
In Section~\ref{sec:SFL-domain}, we present the domain $\SFL$ that
integrates set-sharing, freeness and linearity.
In Section~\ref{sec:SFL-ASub-comparison}, we show that $\SFL$ is uniformly
more precise than the domain $\ASub$, whereas all the previous proposals
for a domain integrating set-sharing and linearity fail to satisfy
such a property.
In Section~\ref{sec:SFL-redundant}, we show that the domain $\SFL$ can be
simplified by removing some redundant
information.
In Section~\ref{sec:exp-eval}, we provide an experimental evaluation
using the \china{} analyzer \cite{Bagnara97th}.
In Section~\ref{sec:related}, we discuss some related work.
Section~\ref{sec:conclusion}
concludes with some final remarks.
The proofs of the results stated here are not included but all of them
are available in an extended version of this paper~\cite{HillBZ03TR}.

\section{Preliminaries}
\label{sec:prelims}

For a set $S$, $\wp(S)$ is the powerset of $S$.
The cardinality of $S$ is denoted by $\card S$ and
the empty set is denoted by $\emptyset$.
The notation $\wpf(S)$ stands for the set of
all the \emph{finite} subsets of $S$,
while the notation $S \sseqf T$ stands for $S \in \wpf(T)$.
The set of all finite sequences of elements of $S$ is denoted
by $S^\ast$, the empty sequence by $\epsilon$, and the concatenation
of $s_1, s_2 \in S^\ast$ is denoted by $s_1 \concat s_2$.

\subsection{Terms and Trees}

Let $\Sig$ denote a possibly infinite set of function symbols,
ranked over the set of natural numbers.
Let $\Vars$ denote a denumerable set of variables, disjoint from $\Sig$.
Then $\Terms$ denotes the free algebra of all (possibly infinite)
terms in the signature $\Sig$ having variables in $\Vars$.
Thus a term can be seen as an ordered labeled tree, possibly having
some infinite paths and possibly containing variables: every inner
node is labeled with a function symbol in $\Sig$ with a rank matching
the number of the node's immediate descendants, whereas every leaf is
labeled by either a variable in $\Vars$ or a function symbol in $\Sig$
having rank $0$ (a constant).
It is assumed that
$\Sig$ contains at least two distinct function
symbols, with one of them having rank~$0$.

If $t \in \Terms$ then $\vars(t)$ and $\mvars(t)$ denote the set and
the multiset of variables occurring in $t$, respectively.
We will also write $\vars(o)$ to denote the set of variables occurring
in an arbitrary syntactic object $o$.

Suppose $s,t \in \Terms$:
$s$ and $t$ are \emph{independent} if
$\vars(s) \inters \vars(t) = \emptyset$;
we say that variable $y$ \emph{occurs linearly in $t$},
more briefly written using the predication $\occlin(y, t)$,
if $y$ occurs exactly once in $\mvars(t)$;
$t$ is said to be \emph{ground} if $\vars(t) = \emptyset$;
$t$ is \emph{free} if $t \in \Vars$;
$t$ is \emph{linear} if, for all $y \in \vars(t)$, we have $\occlin(y, t)$;
finally, $t$ is a \emph{finite term} (or \emph{Herbrand term}) if it
contains a finite number of occurrences of function symbols.
The sets of all ground, linear and finite terms are denoted by
$\GTerms$, $\LTerms$ and $\HTerms$, respectively.

\subsection{Substitutions}

A \emph{substitution} is a total function
$\fund{\sigma}{\Vars}{\HTerms}$ that is the identity almost
everywhere; in other words, the \emph{domain} of~$\sigma$,
\[
  \dom(\sigma)
    \defeq
      \bigl\{\,
        x \in \Vars
      \bigm|
        \sigma(x) \neq x
      \,\bigr\},
\]
is finite.
Given a substitution $\fund{\sigma}{\Vars}{\HTerms}$,
we overload the symbol `$\sigma$' so as to denote also the function
$\fund{\sigma}{\HTerms}{\HTerms}$ defined as follows, for each term
$t \in \HTerms$:
\[
  \sigma(t)
    \defeq
      \begin{cases}
        t,                       &\text{if $t$ is a constant symbol;} \\
        \sigma(t),               &\text{if $t \in \Vars$;} \\
        f\bigl(\sigma(t_1), \ldots, \sigma(t_n)\bigr),
                                 &\text{if $t = f(t_1, \ldots, t_n)$.}
      \end{cases}
\]
If $t \in \HTerms$, we write $t\sigma$ to denote $\sigma(t)$.
Note that, for each substitution $\sigma$
and each finite term $t \in \HTerms$, if $t\sigma \in \Vars$,
then $t \in \Vars$.

If $x \in \Vars$ and $t \in \HTerms \setdiff \{x\}$,
then $x \mapsto t$ is called a \emph{binding}.
The set of all bindings is denoted by $\Bind$.
Substitutions are denoted by the set of their bindings,
thus a substitution $\sigma$ is identified with the (finite) set
\[
  \bigl\{\, x \mapsto x\sigma \bigm| x \in \dom(\sigma) \,\bigr\}.
\]
We denote by $\vars(\sigma)$ the set of variables occurring in the
bindings of $\sigma$.
We also define
\(
  \range(\sigma)
    \defeq
      \bigunion
      \bigl\{\,
        \vars(x\sigma)
      \bigm|
        x \in \dom(\sigma)
      \,\bigr\}
\).

A substitution is said to be \emph{circular} if, for $n > 1$,
it has the form
\[
  \{x_1\mapsto x_2, \ldots, x_{n-1}\mapsto x_n, x_n\mapsto x_1\},
\]
where $x_1$, \ldots, $x_n$ are distinct variables.
A substitution is in \emph{rational solved form} if it has no circular subset.
The set of all substitutions in rational solved form is denoted by
$\RSubst$.
A substitution $\sigma$ is \emph{idempotent} if, for all $t \in \Terms$,
we have $t\sigma\sigma = t\sigma$.
Equivalently, $\sigma$ is idempotent
if and only if $\dom(\sigma) \inters \range(\sigma) = \emptyset$.
The set of all idempotent substitutions
is denoted by $\ISubst$ and $\ISubst \subset \RSubst$.

The composition of substitutions is defined in the usual way.
Thus $\tau \compose \sigma$ is the substitution such that,
for all terms $t \in \HTerms$,
\[
  t(\tau\compose \sigma) = t\sigma\tau
\]
and has the formulation
\begin{equation}
\label{eq:alt-compose}
  \tau\compose \sigma
    =
      \bigl\{\,
        x \mapsto x\sigma\tau
      \bigm|
        x \in \dom(\sigma) \union \dom(\tau),
        x \neq x\sigma\tau
      \,\bigr\}.
\end{equation}
As usual, $\sigma^0$ denotes the identity function
(i.e., the empty substitution)
and, when $i > 0$, $\sigma^i$ denotes the substitution
$(\sigma \compose \sigma^{i-1})$.

For each $\sigma \in \RSubst$ and $s \in \HTerms$, the sequence of finite
terms
\[
  \sigma^0(s), \sigma^1(s), \sigma^2(s), \ldots
\]
converges to a (possibly infinite) term, denoted
$\sigma^\infty(s)$~\cite{IntrigilaVZ96,King00}.
Therefore, the function $\fund{\rt}{\HTerms \times \RSubst}{\Terms}$
such that
\[
  \rt(s, \sigma) \defeq \sigma^\infty(s)
\]
is well defined.
Note that, in general, this function is not a substitution:
while having a finite domain,
its ``bindings'' $x \mapsto \rt(x, \sigma)$ can map a domain variable $x$
into a term $\rt(x, \sigma) \in \Terms \setdiff \HTerms$.
However, as the name of the function suggests,
the term $\rt(x, \sigma)$ is granted to be \emph{rational},
meaning that it can only have a finite number of distinct subterms
and hence, be finitely represented.

\begin{example}
Consider the substitutions
\begin{alignat*}{2}
  \sigma_1
    &= \bigl\{ x \mapsto f(z), y \mapsto a \bigr\}
      &&\in \ISubst, \\
  \sigma_2
    &= \bigl\{ x \mapsto f(y), y \mapsto a \bigr\}
      &&\in \RSubst \setdiff \ISubst, \\
  \sigma_3
    &= \bigl\{x \mapsto f(x) \bigr\}
      &&\in \RSubst \setdiff \ISubst, \\
  \sigma_4
    &= \bigl\{x \mapsto f(y), y \mapsto f(x) \bigr\}
      &&\in \RSubst \setdiff \ISubst, \\
  \sigma_5
    &= \bigl\{x \mapsto y, y \mapsto x \bigr\}
      &&\notin \RSubst.
\end{alignat*}
Note that there are substitutions, such as $\sigma_2$,
that are not idempotent and nonetheless define finite trees only;
namely, $\rt(x, \sigma_2) = f(a)$.
Similarly, there are other substitutions, such as $\sigma_4$,
whose bindings are not explicitly cyclic and nonetheless
define rational trees that are infinite;
namely, $\rt(x, \sigma_4) = f(f(f(\cdots)))$.
Finally note that the `$\rt$' function is not defined
on $\sigma_5 \notin \RSubst$.
\end{example}

\subsection{Equality Theories}

An \emph{equation} is of the form $s = t$ where $s,t \in \HTerms$.
$\Eqs$ denotes the set of all equations.
A substitution $\sigma$
may be regarded as a finite set of equations,
that is, as the set
$\bigl\{\, x = t \bigm| (x \mapsto t) \in \sigma \,\bigr\}$.
We say that a set of equations $e$ is in \emph{rational solved form}
if $\bigl\{\, s \mapsto t \bigm| (s=t) \in e \,\bigr\} \in \RSubst$.
In the rest of the paper, we will often write
a substitution $\sigma \in \RSubst$
to denote a set of equations in rational solved form
(and vice versa).
As is common in research work involving equality,
we overload the symbol `$=$' and use it to denote both equality and to
represent syntactic identity.
The context makes it clear what is intended.

Let $\{r, s, t, s_1, \ldots, s_n, t_1, \ldots, t_n\} \sseq \HTerms$.
We assume that any equality theory $T$ over $\Terms$
includes the \emph{congruence axioms}
denoted by the following schemata:
\begin{align}
\label{eq-ax:id}
s = s &,  \\
\label{eq-ax:sym}
s = t &\piff t = s, \\
\label{eq-ax:trans}
r = s \land s = t &\pimplies r = t, \\
\label{eq-ax:congr}
s_1 = t_1 \land \cdots \land s_n = t_n
  &\pimplies
    f(s_1, \ldots, s_n) = f(t_1, \ldots, t_n).
\end{align}

In logic programming and most implementations of Prolog
it is usual to assume an equality theory based on syntactic identity.
This consists of the congruence axioms
together with the \emph{identity axioms} denoted by
the following schemata,
where $f$ and $g$ are distinct function symbols or $n \neq m$:
\begin{gather}
\label{eq-ax:injective-functions}
f(s_1, \ldots, s_n) = f(t_1, \ldots, t_n)
  \pimplies
    s_1 = t_1 \land \cdots \land s_n = t_n, \\
\label{eq-ax:diff-funct}
\neg \bigl(f(s_1,\ldots,s_n) = g(t_1,\ldots,t_m)\bigr).
\end{gather}
The axioms characterized by
schemata~(\ref{eq-ax:injective-functions}) and~(\ref{eq-ax:diff-funct})
ensure the equality theory depends only on the syntax.
The equality theory for a non-syntactic domain replaces these axioms
by ones that depend instead on the semantics of the domain and,
in particular, on the interpretation given to functor symbols.

The equality theory of Clark~\cite{Clark78}, denoted $\HT$,
on which pure logic programming is based,
usually called the \emph{Herbrand} equality theory,
is given by the congruence axioms, the identity axioms,
and the axiom schema
\begin{equation}
\label{eq-ax:occ-check}
 \forall z \in \Vars \itc
    \forall t \in (\HTerms \setdiff \Vars) \itc
      z \in \vars(t) \pimplies \neg (z = t).
\end{equation}
Axioms characterized by the schema~(\ref{eq-ax:occ-check})
are called the \emph{occurs-check axioms} and are an essential
part of the standard unification procedure in SLD-resolution.

An alternative approach used in some implementations
of logic programming systems,
such as Prolog~II, SICStus and Oz,
does not require the occurs-check axioms.
This approach is based on the theory of
rational trees~\cite{Colmerauer82,Colmerauer84}, denoted $\RT$.
It assumes the congruence axioms and the identity axioms together with a
\emph{uniqueness axiom} for each substitution in rational solved form.
Informally speaking these state that,
after assigning a ground rational tree to each variable
which is not in the domain,
the substitution uniquely defines a ground rational tree
for each of its domain variables.
Note that being in rational solved form is a very weak property.
Indeed, unification algorithms returning a set of equations
in rational solved form are allowed to be much more ``lazy''
than one would expect.
We refer the interested reader to \cite{JaffarLM87,Keisu94th,Maher88}
for details on the subject.

In the sequel we use the expression ``equality theory''
to denote any consistent, decidable theory $T$ satisfying
the congruence axioms.
We also use the expression ``syntactic equality theory''
to denote any equality theory $T$ also satisfying
the identity axioms.

We say that a substitution $\sigma \in \RSubst$ is \emph{satisfiable}
in an equality theory $T$ if, when interpreting $\sigma$ as an
equation system in rational solved form,
\[
  T \entails
      \forall \bigl( \Vars \setdiff \dom(\sigma) \bigr) \itc
        \exists \dom(\sigma) \st
          \sigma.
\]
Let $e \in \wpf(\Eqs)$ be a set of equations in an equality theory $T$.
A substitution $\sigma \in \RSubst$ is called a 
\emph{solution for $e$ in $T$}
if $\sigma$ is satisfiable in $T$ and
$T \entails \forall(\sigma \pimplies e)$;
we say that $e$ is satisfiable if it has a solution.
If $\vars(\sigma) \sseq \vars(e)$, then
$\sigma$ is said to be a \emph{relevant} solution for $e$.
In addition, $\sigma$ is a
\emph{most general solution for $e$ in $T$} if
$T \entails \forall(\sigma \piff e)$.
In this paper, a most general solution is always
a relevant solution of $e$.
When the theory $T$ is clear from the context,
the set of all the relevant most general solutions for
$e$ in $T$ is denoted by $\mgs(e)$.

\begin{example}
\label{ex:equations}
Let
\(
  e = \bigl\{
        g(x) = g(f(y)),
        f(x) = y,
        z = g(w)
      \bigr\}
\)
and
\[
  \sigma
    = \bigl\{
        x \mapsto f(y),
        y \mapsto f(x),
        z \mapsto g(w)
      \bigr\}.
\]
Then, for any syntactic equality theory $T$, we have
\(
  T \entails \forall(\sigma \piff e)
\).
Since $\sigma \in \RSubst$, then $\sigma$ and hence $e$
is satisfiable in $\RT$.
Intuitively, whatever rational tree $t_w$ is assigned to the parameter
variable $w$, there exist rational trees $t_x$, $t_y$ and $t_z$ that,
when assigned to the domain variables $x$, $y$ and $z$,
will turn $\sigma$ into a set of trivial identities;
namely, let $t_x$ and $t_y$ be both equal to
the infinite rational tree $f(f(f(\cdots)))$,
which is usually denoted by $f^\omega$,
and let $t_z$ be the rational tree $g(t_w)$.
Thus $\sigma$ is a relevant most general solution for $e$ in $\RT$.
In contrast,
\[
  \tau
    = \bigl\{
        x \mapsto f(y),
        y \mapsto f(x),
        z \mapsto g(f(a))
      \bigr\}
\]
is just a relevant solution for $e$ in $\RT$.
Also observe that, for any equality theory $T$,
\[
  T \entails \forall \Bigl(
                       \sigma
                         \pimplies
                           \bigl\{ x = f(f(x)) \bigr\}
                     \Bigr)
\]
so that $\sigma$ does not satisfy the occurs-check axioms.
Therefore, neither $\sigma$ nor $e$ are satisfiable
in the Herbrand equality theory $\HT$.
Intuitively, there is no finite tree $t_x$
such that $t_x = f(f(t_x))$.
\end{example}

We have the following useful result regarding `$\rt$' and
satisfiable substitutions that are equivalent with respect to
any given syntactic equality theory.

\begin{proposition}
\label{prop:rt-preserves-Vars-GTerms-LTerms}
Let $\sigma,\tau \in \RSubst$ be satisfiable
in the syntactic equality theory $T$ and
suppose that $T \entails \forall (\sigma \piff \tau)$.
Then 
\begin{align}
\label{prop:rt-preserves-Vars-GTerms-LTerms:Vars}
  \rt(y, \sigma) \in \Vars
    \quad &\iff \quad
      \rt(y, \tau) \in \Vars, \\
\label{prop:rt-preserves-Vars-GTerms-LTerms:GTerms}
  \rt(y, \sigma) \in \GTerms
    \quad &\iff \quad
      \rt(y, \tau) \in \GTerms, \\
\label{prop:rt-preserves-Vars-GTerms-LTerms:LTerms}
  \rt(y, \sigma) \in \LTerms
    \quad &\iff \quad
      \rt(y, \tau) \in \LTerms.
\end{align}
\end{proposition}

\subsection{Galois Connections and Upper Closure Operators}

Given two complete lattices $(C, \leq_C)$ and $(A, \leq_A)$,
a \emph{Galois connection} is a pair of monotonic functions
$\fund{\alpha}{C}{A}$
and
$\fund{\gamma}{A}{C}$
such that
\begin{align*}
  \forall c \in C
    &\itc c \leq_C \gamma\bigl(\alpha(c)\bigr),
  &
  \forall a \in A
    &\itc \alpha\bigl(\gamma(a)\bigr) \leq_A a.
\end{align*}
The functions $\alpha$ and $\gamma$ are said to be the
abstraction and concretization functions, respectively.
A \emph{Galois insertion} is a Galois
connection where the concretization function $\gamma$ is injective.

An \emph{upper closure operator} (uco) $\fund{\rho}{C}{C}$
on the complete lattice $(C, \leq_C)$
is a monotonic, idempotent and extensive\footnote{%
Namely, $c \leq_C \rho(c)$ for each $c \in C$.} self-map.
The set of all uco's on $C$, denoted by $\uco(C)$,
is itself a complete lattice.
For any $\rho \in \uco(C)$,
the set $\rho(C)$, i.e., the image under $\rho$ of the lattice carrier,
is a complete lattice under the same partial order $\mathord{\leq_C}$
defined on $C$.
Given a Galois connection,
the function $\rho \defeq \gamma \compose \alpha$
is an element of $\uco(C)$.
The presentation of abstract interpretation
in terms of Galois connections can be rephrased by using uco's.
In particular, the partial order $\ucoleq$ defined on $\uco(C)$
formalizes the intuition of an abstract domain being
more precise than another one;
moreover, given two elements $\rho_1,\rho_2 \in \uco(C)$,
their reduced product~\cite{CousotC79},
denoted $\rho_1 \ucomeet \rho_2$,
is their $\glb$ on $\uco(C)$.

\subsection{The Set-Sharing Domain}

The set-sharing domain of Jacobs and Langen \cite{JacobsL89},
encodes both aliasing and groundness information.
Let $\VI \sseqf \Vars$ be a fixed and finite set of variables of interest.
An element of the set-sharing domain (a \emph{sharing set}) is
a set of subsets of $\VI$ (the \emph{sharing groups}).
Note that the empty set is not a sharing group.

\begin{definition} \summary{(The \emph{set-sharing} lattice.)}
\label{def:SH}
Let
\(
  \SG \defeq \wp(\VI) \setdiff \{ \emptyset \}
\)
be the set of \emph{sharing groups}.
The set-sharing lattice is defined as
\(
  \SH \defeq \wp(\SG)
\),
ordered by subset inclusion.
\end{definition}

The following operators on $\SH$ are needed for
the specification of the abstract semantics.

\begin{definition} \summary{(Auxiliary operators on $\SH$.)}
\label{def:aux-funcs-SH}
For each $\sh, \sh_1, \sh_2 \in \SH$ and each $V \sseq \VI$,
we define the following functions:

\noindent
the \emph{star-union} function
$\fund{(\cdot)^\star}{\SH}{\SH}$,
is defined as
\begin{align*}
  \sh^\star
    &\defeq
      \bigl\{\,
        S \in \SG
      \bigm|
        \exists n \geq 1
          \st
            \exists S_1, \ldots, S_n \in \sh
              \st S = S_1 \union \cdots \union S_n
      \,\bigr\}; \\
\intertext{%
the extraction of the
\emph{relevant component of $\sh$ with respect to $V$}
is encoded by
$\fund{\rel}{\wp(\VI) \times \SH}{\SH}$
defined as
}
  \rel(V, \sh)
    &\defeq
      \{\, S \in \sh \mid S \inters V \neq \emptyset \,\}; \\
\intertext{%
the \emph{irrelevant component of $\sh$ with respect to $V$} is
thus defined as
}
  \irel(V, \sh)
    &\defeq \sh \setdiff \rel(V, \sh); \\
\intertext{%
the \emph{binary union} function
$\fund{\bin}{\SH \times \SH}{\SH}$
is defined as
}
  \bin(\sh_1, \sh_2)
    &\defeq
      \{\,
        S_1 \union S_2
      \mid
        S_1 \in \sh_1,
        S_2 \in \sh_2
      \,\}; \\
\intertext{%
the \emph{self-bin-union} operation on $\SH$ is defined as
}
\sh^2 
  &\defeq
    \bin(\sh, \sh);\\
\intertext{%
the \emph{abstract existential quantification} function
$\fund{\aproj}{\SH \times \wp(\VI)}{\SH}$
is defined as
}
  \aproj(\sh, V)
    &\defeq
      \bigl\{\,
        S \setdiff V
      \bigm|
        S \in \sh, S \setdiff V \neq \emptyset
      \,\bigr\}
        \union
          \bigl\{\, \{x\} \bigm| x \in V \,\bigr\}.
\end{align*}
\end{definition}

In~\cite{BagnaraHZ97b,BagnaraHZ02TCS} it was shown that the domain $\SH$
contains many elements that are redundant for the computation
of the actual \emph{observable} properties of the analysis,
definite groundness and definite independence.
The following formalization of these observables is a rewording
of the definitions provided in~\cite{ZaffanellaHB99,ZaffanellaHB02TPLP}.       
\begin{definition}\summary{(The observables of $\SH$.)}
\label{def:SH-observables}
The \emph{groundness} and \emph{independence} observables (on $\SH$)
$\rhoCon, \rhoPS \in \uco(\SH)$ are defined,
for each $\sh \in \SH$, by
\begin{align*}
  \rhoCon(\sh)
    &\defeq
      \bigl\{\,
        S \in \SG
      \bigm|
        S \sseq \vars(\sh)
      \,\bigr\}, \\
  \rhoPS(\sh)
    &\defeq
      \bigl\{\,
        S \in \SG
      \bigm|
        (P \sseq S \land \card P = 2)
          \implies
            (\exists T \in \sh \st P \sseq T)
      \,\bigr\}.
\end{align*}
\end{definition}
Note that, as usual in sharing analysis domains, definite groundness and
definite independence are both represented by encoding
possible non-groundness and possible pair-sharing information.

The abstract domain $\PSD$~\cite{BagnaraHZ02TCS,ZaffanellaHB02TPLP}
is the simplest abstraction of the domain $\SH$ that still preserves
the same precision on groundness and independence.
\begin{definition}
\summary{(The \emph{pair-sharing dependency} lattice $\PSD$.)}
\label{def:rhoPSD}
The operator $\rhoPSD \in \uco(\SH)$ is defined,
for each $\sh \in \SH$, by 
\[
  \rhoPSD(\sh)
    \defeq
        \Bigl\{\,
          S \in \SG
        \Bigm|
          \forall y \in S
            \itc 
              S = \bigunion
                    \{\,
                      U \in \sh
                    \mid
                      y \in U \sseq S
                    \,\}
        \,\Bigr\}.
\]
The \emph{pair-sharing dependency} lattice is
\(
  \PSD \defeq  \rhoPSD(\SH)
\).
\end{definition}

In the following example we provide an intuitive interpretation
of the approximation induced by the three upper closure operators
of Definitions~\ref{def:SH-observables} and~\ref{def:rhoPSD}.

\begin{example}
\label{ex:Con-PS}
Let $\VI = \{ v, w, x, y, z \}$ and consider\footnote{%
In this and all the following examples,
we will adopt a simplified notation for a set-sharing element $\sh$,
omitting inner braces.
For instance, we will write $\{xy,xz,yz\}$
to denote $\bigl\{\{x,y\},\{x,z\},\{y,z\}\bigr\}$.
}
$\sh = \{ vx, vy, xy, xyz \}$.
Then
\begin{align*}
  \rhoCon(\sh)
    &= \{
         v, vx, vxy, vxyz, vxz, vy, vyz, vz,
         x, xy, xyz, xz, y, yz, z
       \}, \\
  \rhoPS(\sh)
    &= \{
         v, vx, vxy, vy,
         w,
         x, xy, xyz, xz, y, yz, z
       \}, \\
  \rhoPSD(\sh)
    &= \{ vx, vxy, vy, xy, xyz \}.
\end{align*}
When observing $\rhoCon(\sh)$, the only information available
is that variable $w$ does not occur in a sharing group;
intuitively, this means that $w$ is definitely ground.
All the other information encoded in $\sh$ is lost;
for instance, in $\sh$ variables $v$ and $z$ never occur
in the \emph{same} sharing group
(i.e., they are definitely independent),
while this happens in $\rhoCon(\sh)$.

When observing $\rhoPS(\sh)$, it should be noted that
two distinct variables occur in the same sharing group
if and only if they were also occurring together in
a sharing group of $\sh$,
so that the definite independence information is preserved
(e.g., $v$ and $z$ keep their independence).
On the other hand, all the variables in $\VI$ occur
as singletons
in $\rhoPS(\sh)$ whether or not they are known to be ground;
for instance, $\{w\}$ occurs in $\rhoPS(\sh)$ although $w$ does
not occur in any sharing group in $\sh$.

By noting that $\rhoPSD(\sh) \subset \rhoCon(\sh) \inters \rhoPS(\sh)$,
it follows that $\rhoPSD(\sh)$ preserves both the definite groundness
and the definite independence information of $\sh$;
moreover, as the inclusion is strict, $\rhoPSD(\sh)$ encodes
other information, such as variable covering
(the interested reader is referred
to~\cite{BagnaraHZ02TCS,ZaffanellaHB02TPLP}
for a more formal discussion).
\end{example}

\subsection{Variable-Idempotent Substitutions}

One of the key concepts used in~\cite{HillBZ03TR}
for the proofs of the correctness
results stated in this paper is that of variable-idempotence.
For the interested reader,
we provide here a brief introduction to variable-idempotent substitutions,
although these are not referred to elsewhere in the paper.

The definition of idempotence
requires that repeated applications of a substitution
do not change the syntactic structure of a term  
and idempotent substitutions are normally the preferred form of
a solution to a set of equations.
However, in the domain of rational trees,
a set of solvable equations does not necessarily have an idempotent solution
(for instance, in Example~\ref{ex:equations}, the set of equations
$e$ has no idempotent solution).
On the other hand, several abstractions of terms, such as the ones
commonly used for sharing analysis, are only interested
in the set of variables occurring in a term
and not in the concrete structure that contains them.
Thus, for applications such as sharing analysis,
a useful way to relax the definition of idempotence
is to ignore the structure of terms
and just require that the repeated application of a substitution
leaves the set of variables in a term invariant.

\begin{definition} \summary{(Variable-idempotence.)}
\label{def:VSubst}
A substitution $\sigma \in \RSubst$ is
\emph{vari\-able-idempotent}\footnote{This definition,
which is the same as that originally provided in~\cite{HillBZ98b},
is slightly stronger than the one adopted in~\cite{HillBZ02TPLP},
which disregarded the domain variables of the substitution.
The adoption of this stronger definition allows for some simplifications
in the correctness proofs for freeness and linearity.}
if and only if for all $t \in \HTerms$ we have
\[
  \vars(t\sigma\sigma) = \vars(t\sigma).
\]
The set of variable-idempotent substitutions
is denoted $\VSubst$.
\end{definition}
As any idempotent substitution is also variable-idempotent,
we have $\ISubst \subset \VSubst \subset \RSubst$.

\begin{example}
Consider the following substitutions which are all in $\RSubst$.
\begin{alignat*}{2}
  \sigma_1
    &= \bigl\{ x \mapsto f(y) \bigr\}
      &&\in \ISubst \subset \VSubst, \\
  \sigma_2
    &= \bigl\{ x \mapsto f(x) \bigr\}
      &&\in \VSubst \setdiff \ISubst, \\
  \sigma_3
    &= \bigl\{ x \mapsto f(y,z), y \mapsto f(z,y) \bigr\}
      &&\in \VSubst \setdiff \ISubst, \\
  \sigma_4
    &= \bigl\{ x \mapsto y, y \mapsto f(x,y) \bigr\}
      &&\notin \VSubst.
\end{alignat*}
\end{example}

\section{The Domain $\SFL$}
\label{sec:SFL-domain}

The abstract domain $\SFL$ is made up of three components,
providing different kinds of sharing information regarding
the set of variables of interest $\VI$:
the first component is the set-sharing domain $\SH$
of Jacobs and Langen \cite{JacobsL89};
the other two components provide freeness and linearity information,
each represented by simply recording those variables of interest
that are known to enjoy the corresponding property.

\begin{definition} \summary{(The domain $\SFL$.)}
Let $F \defeq \wp(\VI)$ and $L \defeq \wp(\VI)$
be partially ordered by reverse subset inclusion.
The abstract domain $\SFL$ is defined as
\[
  \SFL
    \defeq
      \bigl\{\,
        \langle \sh, f, l \rangle
      \bigm|
        \sh \in \SH, f \in F, l \in L
      \,\bigr\}
\]
and is ordered by $\mathord{\leqSFL}$,
the component-wise extension of the orderings defined
on the sub-domains.
With this ordering, $\SFL$ is a complete lattice
whose least upper bound operation is denoted by $\alubSFL$.
The bottom element $\langle \emptyset, \VI, \VI \rangle$
will be denoted by $\botSFL$.
\end{definition}

\subsection{The Abstraction Function}
\label{subsec:abstrSFL}

When the concrete domain is based on the theory
of finite trees, idempotent substitutions provide
a finitely computable \emph{strong normal form} for domain elements,
meaning that different substitutions
describe different sets of finite trees.\footnote{%
As usual, this is  modulo the possible renaming
of variables.}
In contrast, when working on a concrete domain based on the theory
of rational trees, substitutions in rational solved form, while being
finitely computable, no longer satisfy this property:
there can be an infinite set of substitutions in rational solved form
all describing the same set of rational trees (i.e., the same
element in the ``intended'' semantics).
For instance, the substitutions
\[
  \sigma_n = \bigl\{ x \mapsto \overbrace{f( \cdots f(}^n x ) \cdots ) \bigr\},
\]
for $n = 1$, $2$,~\dots,
all map the variable $x$ into the same infinite rational tree $f^\omega$.

Ideally, a strong normal form for the set of rational trees described by
a substitution $\sigma \in \RSubst$ can be obtained by computing
the limit $\sigma^\infty$.
The problem is that $\sigma^\infty$ can map domain variables to
infinite rational terms and may not be in $\RSubst$.

This poses a non-trivial problem when trying to define ``good''
abstraction functions, since it would be really desirable for this
function to map any two equivalent concrete elements to the same
abstract element.
As shown in~\cite{HillBZ02TPLP}, the classical abstraction function
for set-sharing analysis~\cite{CortesiF99,JacobsL89}, which was
defined only for substitutions that are idempotent, does not enjoy this
property when applied, as it is, to arbitrary substitutions in
rational solved form.
In~\cite{HillBZ98b,HillBZ02TPLP}, this problem is solved by replacing
the sharing group operator `$\sg$' of~\cite{JacobsL89}
by an occurrence operator, `$\occ$',
defined by means of a fixpoint computation.
However, to simplify the presentation,
here we define `$\occ$' directly by exploiting the fact that
the number of iterations needed to reach the fixpoint
is bounded by the number of bindings in the substitution.

\begin{definition} \summary{(Occurrence operator.)}
\label{def:occ}
For each $\sigma \in \RSubst$ and $v \in \Vars$,
the \emph{occurrence operator}
$\fund{\occ}{\RSubst \times \Vars}{\wpf(\Vars)}$
is defined as
\begin{align*}
  \occ(\sigma, v)
    &\defeq
      \bigl\{\,
        y \in \Vars 
      \bigm|
        n = \card \sigma,  
        v \in \vars(y\sigma^n) \setdiff \dom(\sigma)
      \,\bigr\}. \\
\intertext{%
For each $\sigma \in \RSubst$,
the operator $\fund{\ssets}{\RSubst}{\SH}$ is defined as
}
  \ssets(\sigma)
    &\defeq
      \bigl\{\,
        \occ(\sigma, v) \inters \VI
      \bigm|
        v \in \Vars
      \,\bigr\}
        \setdiff \{ \emptyset \}.
\end{align*}
\end{definition}
The operator `$\ssets$'
is introduced for notational convenience only.

\begin{example}
\label{ex:occ}
Let
\begin{align*}
  \sigma &=
      \bigl\{
        x_1 \mapsto f(x_2),
        x_2 \mapsto g(x_3,x_4),
        x_3 \mapsto x_1
      \bigr\}, \\
  \tau &=
      \bigl\{
        x_1 \mapsto f(g(x_3,x_4)),
        x_2 \mapsto g(x_3,x_4),
        x_3 \mapsto f(g(x_3,x_4))
     \bigr\}.
\end{align*}
Then
$\dom(\sigma) = \dom(\tau) = \{x_1,x_2,x_3\}$
so that
$\occ(\sigma, x_i) = \occ(\tau, x_i) = \emptyset$, for $i = 1$, $2$, $3$ and 
$\occ(\sigma, x_4) = \occ(\tau, x_4) = \{x_1,x_2,x_3,x_4\}$.
As a consequence, supposing that $\VI = \{x_1,x_2,x_3,x_4\}$,
we obtain $\ssets(\sigma) = \ssets(\tau) = \{ \VI \}$.
\end{example}

In a similar way, it is possible to define suitable operators
for groundness, freeness and linearity.
As all ground trees are linear,
a knowledge of the definite groundness information can be useful
for proving properties concerning the linearity abstraction.
Groundness is already encoded in the abstraction for set-sharing
provided in Definition~\ref{def:occ}; nonetheless,
for both a simplified notation and a clearer intuitive reading,
we now explicitly define the set of variables
that are associated to ground trees by a substitution in $\RSubst$.

\pagebreak[4]
\begin{definition} \summary{(Groundness operator.)}
\label{def:gvars}
The \emph{groundness operator} $\fund{\gvars}{\RSubst}{\wpf(\Vars)}$
is defined, for each $\sigma \in \RSubst$,
by
\[
  \gvars(\sigma)
    \defeq
      \bigl\{\,
        y \in \dom(\sigma)
      \bigm|
        \forall v \in \Vars \itc y \notin \occ(\sigma, v)
      \,\bigr\}.
\]
\end{definition}

\begin{example}
\label{ex:gvars}
Consider $\sigma \in \RSubst$ where
\begin{equation*}
  \sigma
     =
      \bigl\{
        x_1 \mapsto x_2,
        x_2 \mapsto f(a),
        x_3 \mapsto x_4,
        x_4 \mapsto f(x_2, x_4)
      \bigr\}.
\end{equation*}
Then $\gvars(\sigma) = \{x_1,x_2,x_3,x_4\}$.
Observe that $x_1 \in \gvars(\sigma)$ although $x_1\sigma \in \Vars$.
Also, $x_3 \in \gvars(\sigma)$
although $\vars(x_3\sigma^i) = \{x_2,x_4\} \neq \emptyset$
for all $i \geq 2$.
\end{example}

As for possible sharing, the definite freeness information
can be extracted from a substitution in rational solved form
by observing the result of a bounded number of applications
of the substitution.

\begin{definition} \summary{(Freeness operator.)}
\label{def:fvars}
The \emph{freeness operator} $\fund{\fvars}{\RSubst}{\wp(\Vars)}$
is defined, for each $\sigma \in \RSubst$,
by
\[
 \fvars(\sigma)
   \defeq  
     \{\,
       y \in \Vars
     \mid
       n = \card \sigma, y \sigma^n \in \Vars
     \,\}.
\]
\end{definition}
As $\sigma \in \RSubst$ has no circular subset, 
$y \in \fvars(\sigma)$ implies
$y\sigma^n \in \Vars \setdiff \dom(\sigma)$.

\begin{example}
\label{ex:fvars}
Let $\VI = \{x_1, x_2, x_3, x_4, x_5\}$ and
consider $\sigma \in \RSubst$ where
\begin{equation*}
  \sigma
    = \bigl\{
        x_1 \mapsto x_2,
        x_2 \mapsto f(x_3),
        x_3 \mapsto x_4,
        x_4 \mapsto x_5
      \bigr\}.
\end{equation*}
Then
$\fvars(\sigma) \inters \VI = \{x_3,x_4,x_5\}$.
Thus $x_1 \notin \fvars(\sigma)$ although $x_1\sigma \in \Vars$.
Also, $x_3 \in \fvars(\sigma)$ although $x_3\sigma \in \dom(\sigma)$.
\end{example}

As in previous cases, the definite linearity information
can be extracted by observing the result of a bounded number
of applications of the considered substitution.

\begin{definition} \summary{(Linearity operator.)}
\label{def:lvars}
The \emph{linearity operator} $\fund{\lvars}{\RSubst}{\wp(\Vars)}$
is defined, for each $\sigma \in \RSubst$,
by
\[
  \lvars(\sigma)
    \defeq
      \bigl\{\,
        y \in \Vars
      \bigm|
        n = \card \sigma,
        \forall z \in \vars(y \sigma^n) \setdiff \dom(\sigma)
          \itc
            \occlin(z, y \sigma^{2n})
      \,\bigr\}.
\]
\end{definition}
In the next example we consider the extraction of linearity
from two substitutions.
The substitution $\sigma$ shows that,
in contrast with the case of set-sharing and freeness,
for linearity we may need to compute up to $2n$ applications,
where $n = \card \sigma$;
the substitution $\tau$ shows that, when observing the term $y\tau^{2n}$,
multiple occurrences of domain variables have to be disregarded.

\begin{example}
\label{ex:lvars}
Let $\VI = \{x_1, x_2, x_3, x_4\}$ and
consider $\sigma \in \RSubst$ where
\[
  \sigma
    = \bigl\{
        x_1 \mapsto x_2,
        x_2 \mapsto x_3,
        x_3 \mapsto f(x_1, x_4)
      \bigr\}.
\]
Then
$\lvars(\sigma) \inters \VI = \{x_4\}$.
Observe that $x_1 \notin \lvars(\sigma)$.
This is because $x_4 \notin \dom(\sigma)$,
$x_1\sigma^3 = f(x_1, x_4)$
so that $x_4 \in \vars(x_1\sigma^3)$
and $x_1 \sigma^6 = f\bigl(f(x_1, x_4), x_4\bigr)$
so that $\occlin(x_4, x_1\sigma^6)$ does not hold.
Note also that $\occlin(x_4, x_1\sigma^i)$ holds for $i = 3$, $4$,~$5$.

Consider now $\tau \in \RSubst$ where
\[
  \tau
    = \bigl\{
        x_1 \mapsto f(x_2, x_2),
        x_2 \mapsto f(x_2)
      \bigr\}.
\]
Then $\lvars(\tau) \inters \VI = \VI$.
Note that we have $x_1 \in \lvars(\tau)$
although, for all $i > 0$,
$x_2 \in \dom(\tau)$ occurs more than once
in the term $x_1\tau^i$.
\end{example}

The occurrence, groundness, freeness and linearity operators
are invariant with respect to substitutions that are equivalent
in the given syntactic equality theory.
\begin{proposition}
\label{prop:iff-RSubst-occ-f-g-l}
Let $\sigma, \tau \in \RSubst$ be satisfiable
in the syntactic equality theory $T$ and
suppose that $T \entails \forall(\sigma \piff \tau)$.
Then
\begin{align}
\label{prop:iff-RSubst-occ}
\ssets(\sigma) &= \ssets(\tau), \\
\label{prop:iff-RSubst-g}
\gvars(\sigma) &= \gvars(\tau), \\
\label{prop:iff-RSubst-f}
\fvars(\sigma) &= \fvars(\tau), \\
\label{prop:iff-RSubst-l}
\lvars(\sigma) &= \lvars(\tau).
\end{align}
\end{proposition}

Moreover, these operators precisely capture the intended properties
over the domain of rational trees.
\begin{proposition}
\label{prop:occ-f-g-l-rt-RSubst}
If $\sigma \in \RSubst$ and $y,v \in \Vars$ then
\begin{align}
\label{case:RSubst-occ}
  y \in \occ(\sigma, v)
    \quad &\iff \quad
      v \in \vars\bigl(\rt(y, \sigma)\bigr), \\
\label{case:RSubst-ground}
  y \in \gvars(\sigma)
    \quad &\iff \quad
      \rt(y, \sigma) \in \GTerms, \\
\label{case:RSubst-free}
  y \in \fvars(\sigma)
    \quad &\iff \quad
      \rt(y, \sigma) \in \Vars, \\
\label{case:RSubst-linear}
  y \in \lvars(\sigma)
    \quad &\iff \quad
      \rt(y, \sigma) \in \LTerms.
\end{align}
\end{proposition}
It follows from (\ref{case:RSubst-occ}) and (\ref{case:RSubst-free})
that any free variable necessarily shares
(at least, with itself).
Also, as $\Vars \union \GTerms \sslt \LTerms$,
it follows from (\ref{case:RSubst-ground}), (\ref{case:RSubst-free})
and (\ref{case:RSubst-linear})
that any variable that is
either ground or free is also necessarily linear.
Thus we have the following corollary.
\begin{corollary}
\label{cor:f-g-l-RSubst}
If $\sigma \in \RSubst$,
then 
\begin{align*}
\fvars(\sigma) &\sseq \vars\bigl(\ssets(\sigma)\bigr),\\
\fvars(\sigma) \union \gvars(\sigma) &\sseq \lvars(\sigma).
\end{align*}
\end{corollary}

We are now in position to define the abstraction function
mapping rational trees to elements of the domain $\SFL$.
\begin{definition}\label{def:abstrSFL}
\summary{(The abstraction function for $\SFL$.)}
For each substitution $\sigma \in \RSubst$,
the function $\fund{\abstrSFL}{\RSubst}{\SFL}$ is defined by
\begin{align*}
  \abstrSFL(\sigma)
    &\defeq
       \bigl\langle
         \ssets(\sigma),
         \fvars(\sigma) \inters \VI,
         \lvars(\sigma) \inters \VI
       \bigr\rangle,
\intertext{%
The concrete domain $\wp(\RSubst)$ is related to $\SFL$
by means of the \emph{abstraction function}
$\fund{\abstrSFL}{\wp(\RSubst)}{\SFL}$
such that, for each $\Sigma \in \wp(\RSubst)$,
}
  \abstrSFL(\Sigma)
    &\defeq
      \alubSFL
        \bigl\{\,
          \abstrSFL(\sigma)
        \bigm|
          \sigma \in \Sigma
        \,\bigr\}. \\
\intertext{%
Since the abstraction function $\abstrSFL$ is additive,
the concretization function is given by the adjoint~\textup{\cite{CousotC77}}
}
  \concrSFL\bigl( \langle \sh, f, l \rangle \bigr)
    &\defeq
      \bigl\{\,
        \sigma \in \RSubst
      \bigm|
        \ssets(\sigma) \sseq \sh,
        \fvars(\sigma) \Sseq f,
        \lvars(\sigma) \Sseq l
      \,\bigr\}.
\end{align*}
\end{definition}

With Definition~\ref{def:abstrSFL}
and Proposition~\ref{prop:iff-RSubst-occ-f-g-l},
one of our objectives is fulfilled:
substitutions in $\RSubst$ that are equivalent have the same abstraction.
\begin{corollary}
\label{cor:iff-RSubst-same-abstrSFL}
Let $\sigma, \tau \in \RSubst$ be satisfiable
in the syntactic equality theory $T$ and suppose
$T \entails \forall(\sigma \piff \tau)$.
Then $\abstrSFL(\sigma) = \abstrSFL(\tau)$.
\end{corollary}
Observe that the Galois connection defined by
the functions $\abstrSFL$ and $\concrSFL$ is not a Galois insertion
since different abstract elements are mapped by $\concrSFL$
to the same set of concrete computation states.
To see this it is sufficient to observe that,
by Corollary~\ref{cor:f-g-l-RSubst},
any abstract element $d = \langle \sh, f, l \rangle \in \SFL$
such that $f \Nsseq \vars(\sh)$,
as is the case for the bottom element $\botSFL$,
satisfies
\(
  \concrSFL(d) = \concrSFL(\botSFL) = \emptyset
\);
thus, all such $d$'s will represent the semantics of those program fragments
that have no successful computations.
Similarly, by letting $V = \bigl(\VI \setdiff \vars(\sh)\bigr) \union f$,
it can be seen that, for any $l'$
such that
\(
   V \union l = V \union l',
\)
we have, again by Corollary~\ref{cor:f-g-l-RSubst},
\(
   \concrSFL(d) = \concrSFL\bigl(\langle \sh, f, l' \rangle\bigr)
\).

Of course, by taking the abstract domain as the subset of $\SFL$ that is
the co-domain of $\abstrSFL$, we would have a Galois insertion.
However, apart from the simple cases shown above,
it is somehow difficult
to \emph{explicitly} characterize such a set.
For instance, as observed in~\cite{File94}, if
\begin{equation*}
  \sfl = \bigl\langle
    \{xy, xz, yz\},
    \{x,y,z\},
    \{x,y,z\}
  \bigr\rangle
    \in \SFL
\end{equation*}
we have $\concrSFL(\sfl) = \concrSFL(\botSFL) = \emptyset$.
It is worth stressing that these ``spurious'' elements
do not compromise the correctness of the analysis
and, although they can affect the precision of the analysis,
they rarely occur in practice~\cite{BagnaraZH00,Zaffanella01th}.

\subsection{The Abstract Operators}
\label{subsec:SFL-operators}

The specification of the abstract unification operator on the domain $\SFL$
is rather complex, since it is based on a very detailed case analysis.
To achieve some modularity, that will be also useful when proving
its correctness, in the next definition we introduce
several auxiliary abstract operators.

\begin{definition} \summary{(Auxiliary operators in $\SFL$.)}
\label{def:aux-funcs-SFL}
Let $s,t \in \HTerms$ be finite terms such that
$\vars(s) \union \vars(t) \sseq \VI$.
For each $\sfl = \langle \sh, f, l \rangle \in \SFL$
we define the following predicates:

\noindent
$s$ and $t$ are \emph{independent in $\sfl$}
if and only if
$\fund{\ind_{\sfl}}{\HTerms^2}{\Bool}$
holds for $(s, t)$, where
\begin{align*}
  \ind_\sfl(s, t)
    &\defeq
      \Bigl(
        \rel\bigl(\vars(s), \sh\bigr)
          \inters
            \rel\bigl(\vars(t), \sh\bigr)
              = \emptyset
      \Bigr); \\
\intertext{%
$t$ is \emph{ground in $\sfl$}
if and only if
$\fund{\ground_{\sfl}}{\HTerms}{\Bool}$
holds for $t$, where
}
  \ground_\sfl(t)
    &\defeq
      \bigl(\vars(t) \sseq \VI \setdiff \vars(\sh) \bigr); \\
\intertext{%
$y \in \vars(t)$ \emph{occurs linearly (in $t$) in $\sfl$}
if and only if
$\fund{\occlin_{\sfl}}{\VI \times \HTerms}{\Bool}$
holds for $(y, t)$, where
}
  \occlin_\sfl(y,t)
    &\defeq
      \ground_\sfl(y)
        \lor
          \Bigl(
            \occlin(y, t)
              \land
            (y \in l) \\
    &\qquad \qquad
	      \land
            \forall z \in \vars(t)
              \itc
                \bigl(
                  y \neq z \implies \ind_\sfl(y,z)
                \bigr)
           \Bigr); \\
\intertext{%
$t$ is \emph{free in $\sfl$}
if and only if
$\fund{\free_{\sfl}}{\HTerms}{\Bool}$
holds for $t$, where
}
  \free_\sfl(t)
    &\defeq
      (t \in f); \\
\intertext{%
$t$ is \emph{linear in $\sfl$}
if and only if
$\fund{\lin_{\sfl}}{\HTerms}{\Bool}$
holds for $t$, where
}
  \lin_\sfl(t)
    &\defeq
      \forall y \in \vars(t)
        \itc
          \occlin_\sfl(y, t).
\end{align*}

The function $\fund{\sharewith_{\sfl}}{\HTerms}{\wp(\VI)}$
yields the set of variables of interest
that may share with the given term.
For each $t \in \HTerms$, 
\begin{equation*}
  \sharewith_\sfl(t)
    \defeq
      \vars\Bigl( \rel\bigl(\vars(t), \sh\bigr) \Bigr).
\end{equation*}

The function $\fund{\cyclicreduce_x^t}{\SH}{\SH}$
strengthens the sharing set $\sh$ by forcing the coupling
of $x$ with $t$.
For each $\sh \in \SH$ and each $(x \mapsto t) \in \Bind$, 
\begin{equation*}
  \cyclicreduce_x^t(\sh)
    \defeq
      \irel\bigl(\{x\} \union \vars(t), \sh\bigr)
        \union
          \rel\bigl(\vars(t) \setdiff \{x\}, \sh\bigr).
\end{equation*}
\end{definition}

As a first correctness result, we have that the auxiliary
operators correctly approximate the corresponding concrete properties.
\begin{theorem}
\label{thm:soundness-of-sfl-preds}
Let $\sfl \in \SFL$, $\sigma \in \concrSFL(\sfl)$
and $y \in \VI$.
Let also $s,t \in \HTerms$ be two finite terms such that
$\vars(s) \union \vars(t) \sseq \VI$.
Then
\begin{align}
\label{thm:soundness-of-sfl-preds:ind}
  \ind_\sfl(s, t)
    &\implies
      \vars\bigl(\rt(s, \sigma)\bigr)
        \inters
          \vars\bigl(\rt(t, \sigma)\bigr)
            = \emptyset; \\
\label{thm:soundness-of-sfl-preds:sharewith}
  \ind_\sfl(y, t)
    &\iff
      y \notin \sharewith_{\sfl}(t); \\
\label{thm:soundness-of-sfl-preds:free}
  \free_{\sfl}(t)
    &\implies
      \rt(t, \sigma) \in \Vars; \\
\label{thm:soundness-of-sfl-preds:ground}
  \ground_{\sfl}(t)
    &\implies
      \rt(t, \sigma) \in \GTerms; \\
\label{thm:soundness-of-sfl-preds:lin}
  \lin_{\sfl}(t)
    &\implies
      \rt(t, \sigma) \in \LTerms.
\end{align}
\end{theorem}

\begin{example}
Let $\VI = \{ v, w, x, y, z \}$ and consider the abstract element
\(
  \sfl = \langle \sh, f, l \rangle \in \SFL
\),
where
\begin{align*}
  \sh &= \{ v, wz, xz, z \},
&   f &= \{ v \},
&   l &= \{ v, x, y, z \}.
\end{align*}
Then, by applying Definition~\ref{def:aux-funcs-SFL},
we obtain the following.
\begin{itemize}
\item
$\ground_\sfl(x)$ does not hold whereas
$\ground_\sfl\bigl( h(y) \bigr)$ holds.
\item
$\free_\sfl(v)$ holds but
$\free_\sfl\bigl( h(v) \bigr)$ does not hold.
\item
Both $\ind_\sfl(w, x)$ and
$\ind_\sfl\bigl( f(w,y), f(x,y) \bigr)$ hold
whereas $\ind_\sfl(x, z)$ does not hold;
note that, in the second case, the two arguments of the predicate
do share $y$, but this does not affect the independence
of the corresponding terms, because $y$ is definitely ground
in the abstract element $\sfl$.
\item
Let $t = f(w,x,x,y,y,z)$; then
$\occlin_\sfl(w, t)$ does not hold because $w \notin l$;
$\occlin_\sfl(x, t)$ does not hold because $x$ occurs
more than once in $t$;
$\occlin_\sfl(y, t)$ holds, even though $y$ occurs
twice in $t$, because $y$ is definitely ground in $\sfl$;
$\occlin_\sfl(z, t)$ does not hold because both $x$ and $z$
occur in term $t$ and, as observed in the point above,
$\ind_\sfl(x, z)$ does not hold.
\item
For the reasons given in the point above,
$\lin_\sfl(t)$ does not hold;
in contrast,
$\lin_\sfl\bigl(f(y,y,z)\bigr)$ holds.
\item
$\sharewith_\sfl(w) = \{w, z\}$ and $\sharewith_\sfl(x) = \{x, z\}$;
thus, both $w$ and $x$ may share one or more variables with $z$;
since we observed that $w$ and $x$ are definitely independent in $\sfl$,
this means that the set of variables that $w$ shares with $z$ is disjoint from
the set of variables that $x$ shares with $z$.
\item
Let $t = f(w,z)$; then
\begin{align*}
  \cyclicreduce_z^t(\sh)
    &= \irel\bigl(\{w,z\}, \sh\bigr)
         \union
           \rel\bigl(\{w\}, \sh\bigr) \\
    &= \{ v \} \union \{ wz \} \\
    &= \sh \setdiff \{ xz, z \}.
\end{align*}
An intuitive explanation of the usefulness of this operator
is deferred until after the introduction of the abstract $\mgu$ operator
(see also Example~\ref{ex:amguSFL-cyclicreduce}).
\end{itemize}
\end{example}

We now introduce the abstract $\mgu$ operator,
specifying how a single binding affects each component of
the domain $\SFL$ in the context of a syntactic equality theory $T$.

\begin{definition} \summary{($\amguSFL$.)}
\label{def:amguSFL}
The function $\fund{\amguSFL}{\SFL\times\Bind}{\SFL}$
captures the effects of a binding on an element of $\SFL$.
Let $\sfl = \langle \sh, f, l \rangle \in \SFL$ and
$(x \mapsto t) \in \Bind$, where $\{x\} \union \vars(t) \sseq \VI$.
Let also
\[
  \sh'
    \defeq
      \cyclicreduce_x^t(\sh_{-} \union \sh''),
\]
where
\begin{align*}
  \sh_x
    &\defeq
      \rel\bigl(\{x\}, \sh\bigr),
& \sh_t
    &\defeq
      \rel\bigl(\vars(t), \sh\bigr), \\
  \sh_{xt}
    &\defeq
      \sh_x \inters \sh_t,
& \sh_{-}
    &\defeq
      \irel\bigl(\{x\} \union \vars(t), \sh\bigr),
\end{align*}
\begin{equation*}
  \sh''
    \defeq
      \begin{cases}
        \bin(\sh_x, \sh_t),
          &\text{if $\free_{\sfl}(x)
                       \lor
                     \free_{\sfl}(t)$;} \\
        \bin\bigl(
              \sh_x \union \bin(\sh_x, \sh_{xt}^\star),
          & \\
        \qquad\qquad
              \sh_t \union \bin(\sh_t, \sh_{xt}^\star)
            \bigr),
          &\text{if $\lin_{\sfl}(x) \land \lin_{\sfl}(t)$;} \\
        \bin(\sh_x^\star, \sh_t),
          &\text{if $\lin_{\sfl}(x)$;} \\
        \bin(\sh_x, \sh_t^\star),
          &\text{if $\lin_{\sfl}(t)$;} \\
        \bin(\sh_x^\star, \sh_t^\star),
          &\text{otherwise.}
      \end{cases}
\end{equation*}

Letting $S_x \defeq \sharewith_\sfl(x)$
and $S_t \defeq \sharewith_\sfl(t)$,
we also define
\begin{align*}
  f'
    &\defeq
      \begin{cases}
        f,
          &\text{if $\free_{\sfl}(x) \land \free_{\sfl}(t)$;} \\
        f \setdiff S_x,
          &\text{if $\free_{\sfl}(x)$;} \\
        f \setdiff S_t,
          &\text{if $\free_{\sfl}(t)$;} \\
        f \setdiff (S_x \union S_t),
          &\text{otherwise;} \\
      \end{cases} \\
\displaybreak[0]
  l'
    &\defeq
      \bigl( \VI \setdiff \vars(\sh') \bigr) \union f' \union l'', \\
\intertext{%
where
}
  l''
    &\defeq
      \begin{cases}
        l \setdiff (S_x \inters S_t),
          &\text{if $\lin_{\sfl}(x)
                       \land
                     \lin_{\sfl}(t)$;} \\
        l \setdiff S_x,
          &\text{if $\lin_{\sfl}(x)$;} \\
        l \setdiff S_t,
          &\text{if $\lin_{\sfl}(t)$;} \\
        l \setdiff (S_x \union S_t),
          &\text{otherwise.} \\
      \end{cases}
\intertext{%
Then
}
  \amguSFL\bigl(\sfl, x \mapsto t\bigr)
    &\defeq
      \begin{cases}
        \botSFL,
          &\text{if \( \sfl = \botSFL
                         \lor
                       \bigl(T = \HT \land x \in \vars(t)\bigr)
                    \);} \\
        \langle \sh', f', l' \rangle
          &\text{otherwise.}
      \end{cases}
\end{align*}
\end{definition}

The next result states that the abstract $\mgu$ operator
is a correct approximation of the concrete one.
\begin{theorem}
\label{thm:soundness-of-amguSFL}
Let $\sfl \in \SFL$ and $(x \mapsto t) \in \Bind$,
where $\{x\} \union \vars(t) \sseq \VI$.
Then, for all $\sigma \in \concrSFL(\sfl)$ and
$\tau \in \mgs\bigl(\sigma \union \{x = t\}\bigr)$
in the syntactic equality theory $T$, we have
\(
  \tau \in \concrSFL\bigl(\amguSFL(\sfl, x \mapsto t)\bigr).
\)
\end{theorem}

We now highlight the similarities and differences
of the operator $\amguSFL$ with respect to the corresponding
ones defined in the ``classical'' proposals for the integration
of set-sharing with freeness and linearity,
such as~\cite{BruynoogheCM94,BruynoogheCM95,HansW92,Langen90th}.
Note that, when comparing our domain with the proposal
in~\cite{BruynoogheCM94}, we deliberately ignore
all those enhancements that depend on properties that cannot be
represented in $\SFL$ (i.e., compoundness and
explicit structural information).
\begin{itemize}
\item
In the computation of the set-sharing component,
the main difference can be observed in the second, third and fourth cases
of the definition of $\sh''$:
here we omit one of the star-unions
even when the terms $x$ and $t$ possibly share.
In contrast, in~\cite{BruynoogheCM94,HansW92,Langen90th}
the corresponding star-union is avoided
only when $\ind_\sfl(x, t)$ holds.
Note that when $\ind_\sfl(x, t)$ holds in the second case of $\sh''$,
then we have $\sh_{xt} = \emptyset$;
thus, the whole computation for this case reduces
to $\sh'' = \bin(\sh_x, \sh_t)$,
as was the case in the previous proposals.
\item
Another improvement on the set-sharing component
can be observed in the definition of $\sh'$:
the $\cyclicreduce_x^t$ operator allows the set-sharing description
to be further enhanced when dealing with \emph{explicitly cyclic bindings},
i.e., when $x \in \vars(t)$.
This is the rewording of a similar enhancement
proposed in~\cite{Bagnara97th} for the domain $\Pos$
in the context of groundness analysis.
Its net effect is to recover some groundness and sharing dependencies
that would have been unnecessarily lost when using the standard operators.
When $x \notin \vars(t)$, we have
\(
  \cyclicreduce_x^t(\sh_{-} \union \sh'') = \sh_{-} \union \sh''
\).
\item
The computation of the freeness component $f'$ is the same as
specified in~\cite{BruynoogheCM94,HansW92},
and is more precise than the one defined in~\cite{Langen90th}. 
\item
The computation of the linearity component $l'$ is the same as
specified in~\cite{BruynoogheCM94}, and is more precise than
those defined in~\cite{HansW92,Langen90th}.
\end{itemize}

In the following examples
we show that the improvements in the abstract computation
of the sharing component allow, in particular cases,
to derive better information than that obtainable
by using the classical abstract unification operators.

\begin{example}
\label{ex:amguSFL-both-lin-share}
Let $\VI = \{x, x_1, x_2, y, y_1, y_2, z\}$
and $\sigma \in \RSubst$ such that
\[
  \sigma
    \defeq
      \bigl\{
        x \mapsto f(x_1, x_2, z),
        y \mapsto f(y_1, z, y_2)
      \bigr\}.
\]
By Definition~\ref{def:abstrSFL}, we have
\(
  \sfl
    \defeq \abstrSFL\bigl( \{\sigma\} \bigr)
    = \langle \sh, f, l \rangle
\),
where
\begin{align*}
  \sh &= \{ xx_1, xx_2, xyz, yy_1, yy_2 \},
& f   &= \VI \setdiff \{x,y\},
& l   &= \VI.
\end{align*}
Consider the binding $(x \mapsto y) \in \Bind$.
In the concrete domain, we compute
(a substitution equivalent to)
$\tau \in \mgs\bigl( \sigma \union \{ x=y \} \bigr)$,
where
\[
  \tau = \bigl\{
           x   \mapsto f(y_1, y_2, y_2),
           y   \mapsto f(y_1, y_2, y_2),
           x_1 \mapsto y_1, 
           x_2 \mapsto y_2, 
           z   \mapsto y_2
         \bigr\}.
\]
Note that
\(
  \abstrSFL\bigl( \{\tau\} \bigr)
    = \langle \sh_\tau, f_\tau, l_\tau \rangle
\),
where $\sh_\tau = \{ xx_1yy_1, xx_2yy_2z \}$,
so that the pairs of variables
$P_x = \{x_1, x_2\}$ and
$P_y = \{y_1, y_2\}$ keep their independence.

When evaluating the sharing component of $\amguSFL(d, x \mapsto y)$,
using the notation of Definition~\ref{def:amguSFL},
we have
\begin{align*}
  \sh_x
     &= \{ xx_1, xx_2, xyz \},
&   \sh_t
     &= \{ xyz, yy_1, yy_2 \}, \\
  \sh_{xt}
     &= \{ xyz \},
& \sh_{-}
     &= \emptyset.
\end{align*}
Since both $\lin_\sfl(x)$ and $\lin_\sfl(y)$ hold,
we apply the second case of the definition of $\sh''$ so that
\begin{align*}
  \sh_x \union \bin(\sh_x, \sh_{xt}^\star)
    &= \{ xx_1, xx_1yz, xx_2, xx_2yz, xyz \}, \\
  \sh_t \union \bin(\sh_t, \sh_{xt}^\star)
    &= \{ xyy_1z, xyy_2z, xyz, yy_1, yy_2 \}, \\
  \sh''
    &= \bin\bigl(
             \sh_x \union \bin(\sh_x, \sh_{xt}^\star),
             \sh_t \union \bin(\sh_t, \sh_{xt}^\star)
           \bigr) \\
    &= \{ xx_1yy_1, xx_1yy_1z, xx_1yy_2, xx_1yy_2z, xx_1yz, \\
    &\quad \quad
         xx_2yy_1, xx_2yy_1z, xx_2yy_2, xx_2yy_2z, xx_2yz, \\
    &\quad \quad
         xyy_1z, xyy_2z, xyz \}.
\end{align*}
Finally, as the binding is not cyclic, we obtain $\sh' = \sh''$.
Thus $\amguSFL$ captures the fact that pairs $P_x$ and $P_y$
keep their independence.

In contrast,
since $\ind_\sfl(x, y)$ does not hold,
all of the classical definitions of abstract unification
would have required the star-closure of both $\sh_x$ and $\sh_t$,
resulting in an abstract element including, among the others,
the sharing group $S = \{ x, x_1, x_2, y, y_1, y_2 \}$.
Since $P_x \union P_y \subset S$,
this independence information would have been unnecessarily lost.
\end{example}

Similar examples can be devised for the third and fourth cases
of the definition of $\sh''$, where only one side of the binding
is known to be linear.
The next example shows the precision improvements arising
from the use of the $\cyclicreduce_x^t$ operator.

\begin{example}
\label{ex:amguSFL-cyclicreduce}
Let $\VI = \{x, x_1, x_2, y\}$
and
\(
  \sigma
    \defeq
      \bigl\{
        x \mapsto f(x_1, x_2)
      \bigr\}
\).
By Definition~\ref{def:abstrSFL}, we have
\(
  \sfl
    \defeq \abstrSFL\bigl( \{\sigma\} \bigr)
    = \langle \sh, f, l \rangle
\),
where
\begin{align*}
  \sh &= \{ xx_1, xx_2, y \},
& f   &= \VI \setdiff \{x\},
& l   &= \VI.
\end{align*}
Let $t = f(x,y)$ and consider the cyclic binding
$(x \mapsto t) \in \Bind$.
In the concrete domain, we compute (a substitution equivalent to)
$\tau \in \mgs\bigl( \sigma \union \{ x = t \} \bigr)$,
where
\begin{equation*}
  \tau = \bigl\{
           x   \mapsto f(x_1, x_2),
           x_1 \mapsto f(x_1, x_2),
           y   \mapsto x_2,
         \bigr\}.
\end{equation*}
Note that if we further instantiate $\tau$ by grounding $y$,
then variables $x$, $x_1$ and $x_2$ would become ground too.
Formally we have 
\(
  \abstrSFL\bigl( \{\tau\} \bigr)
    = \langle \sh_\tau, f_\tau, l_\tau \rangle
\),
where $\sh_\tau = \{ xx_1x_2y \}$.
Thus, as observed above, $y$ covers $x$, $x_1$ and $x_2$.
When abstractly evaluating the binding, we compute
\begin{align*}
  \sh_x
     &= \{ xx_1, xx_2 \},
& \sh_t
     &= \{ xx_1, xx_2, y \}, \\
  \sh_{xt}
     &= \sh_x,
& \sh_{-}
     &= \emptyset.
\end{align*}
Since both $\lin_\sfl(x)$ and $\lin_\sfl(t)$ hold,
we apply the second case of the definition of $\sh''$, so that
\begin{align*}
  \sh_x \union \bin(\sh_x, \sh_{xt}^\star)
    &= \sh_x^\star
     = \{ xx_1, xx_1x_2, xx_2 \}, \\
  \sh_t \union \bin(\sh_t, \sh_{xt}^\star)
    &= \{ xx_1, xx_1x_2, xx_1x_2y, xx_1y, xx_2, xx_2y, y \}, \\
  \sh''
    &= \bin\bigl(
             \sh_x \union \bin(\sh_x, \sh_{xt}^\star),
             \sh_t \union \bin(\sh_t, \sh_{xt}^\star)
           \bigr) \\
    &= \{ xx_1, xx_1x_2, xx_1x_2y, xx_1y, xx_2, xx_2y \}.
\end{align*}
Thus, as $x \in \vars(t)$, we obtain
\begin{align*}
  \sh'
     &= \cyclicreduce_x^t(\sh_{-} \union \sh'') \\
     &= \irel\bigl(\{x\} \union \vars(t), \sh''\bigr)
          \union
            \rel\bigl(\vars(t) \setdiff \{x\}, \sh''\bigr) \\
     &= \emptyset \union \rel\bigl(\{y\}, \sh''\bigr) \\
     &= \{ xx_1x_2y, xx_1y, xx_2y \}.
\end{align*}
Note that, in the element $\sh_{-} \union \sh'' = \sh''$
(which is the abstract element that would have been computed
when not exploiting the $\cyclicreduce_x^t$ operator)
variable $y$ covers none of variables $x$, $x_1$ and $x_2$.
Thus, by applying the $\cyclicreduce_x^t$ operator,
this covering information is restored.
\end{example}

The full abstract unification operator $\aunifySFL$,
capturing the effect of a sequence of bindings on an abstract element,
can now be specified by a straightforward inductive definition
using the operator $\amguSFL$.

\begin{definition} \summary{($\aunifySFL$.)}
\label{def:aunifySFL}
The operator $\fund{\aunifySFL}{\SFL \times \Bind^\ast}{\SFL}$ is defined,
for each $\sfl \in \SFL$
and each sequence of bindings $\bs \in \Bind^\ast$, by
\[
  \aunifySFL(\sfl, \bs)
    \defeq
      \begin{cases}
        \sfl,
          &\text{if $\bs = \epsilon$;} \\
        \aunifySFL\bigl( \amguSFL(\sfl, x \mapsto t), \bs'\bigr),
          &\text{if $\bs = (x \mapsto t) \concat \bs'$.}
      \end{cases}
\]
\end{definition}
Note that the second argument of $\aunifySFL$
is a \emph{sequence} of bindings
(i.e., it is not a substitution, which is a \emph{set} of bindings),
because $\amguSFL$ is neither commutative nor idempotent,
so that the multiplicity and the actual order of application
of the bindings can influence the overall result of
the abstract computation.
The correctness of the $\aunifySFL$ operator
is simply inherited from the correctness of the underlying
$\amguSFL$ operator.
In particular, any reordering of the bindings in the sequence $\bs$
still results in a correct implementation of $\aunifySFL$.

The `merge-over-all-path' operator on the domain $\SFL$
is provided by $\alubSFL$ and is correct by definition.
Finally, we define the abstract existential quantification operator
for the domain $\SFL$, whose correctness does not pose any problem.

\begin{definition}\summary{($\aprojSFL$.)}
\label{def:aprojSFL}
The function $\fund{\aprojSFL}{\SFL \times \wpf(\VI)}{\SFL}$
provides the \emph{abstract existential quantification}
of an element with respect to a subset of the variables of interest.
For each $\sfl \defeq \langle \sh, f, l \rangle \in \SFL$
and $V \sseq \VI$,
\begin{align*}
  \aprojSFL\bigl(\langle \sh, f, l \rangle, V\bigr)
    &\defeq
      \bigl\langle
        \aproj(\sh, V), f \union V, l \union V
      \bigr\rangle.
\end{align*}
\end{definition}

The intuition behind the definition of
the abstract operator $\aprojSFL$ is the following.
As explained in Section~\ref{sec:prelims},
any substitution $\sigma \in \RSubst$ can be interpreted,
under the given equality theory $T$, as a first-order logical formula;
thus, for each set of variables $V$, it is possible to consider
the (concrete) existential quantification $\exists V \st \sigma$.
The goal of the abstract operator $\aprojSFL$ is to provide
a correct approximation of such a quantification
starting from any correct approximation for $\sigma$.

\begin{example}
Let $\VI = \{ x, y, z \}$ and
\(
  \sigma
    = \{
        x \mapsto f(v_1, v_2),
        y \mapsto g(v_2, v_3),
        z \mapsto f(v_1, v_1)
      \}
\),
so that, by Definition~\ref{def:abstrSFL},
\[
  \sfl = \abstrSFL\bigl( \{ \sigma \} \bigr)
       = \bigl\langle
           \{ xy, xz, y \},
           \emptyset,
           \{ x, y \}
         \bigr\rangle.
\]
Let $V = \{ y,z \}$ and consider the concrete element corresponding
to the logical formula $\exists V \st \sigma$.
Note that
\(
  T \entails \forall (\tau \piff \exists V \st \sigma)
\),
where
\(
  \tau = \{ x \mapsto f(v_1, v_2) \}
\).
By applying Definition~\ref{def:aprojSFL}, we obtain
\[
  \aprojSFL(\sfl, V)
    = \bigl\langle
        \{ x, y, z \},
        \{ y, z \},
        \{ x, y, z \}
     \bigr\rangle
    = \abstrSFL\bigl( \{ \tau \} \bigr).
\]
It is worth stressing that such an operator does not affect
the set $\VI$ of the variables of interest.
In particular, the abstract element $\aprojSFL(\sfl, V)$ still has to
provide correct information about variables $y$ and $z$.
Intuitively, since all the occurrences of $y$ and $z$
in $\exists V \st \sigma$ are bound by the existential quantifier,
the two variables of interest are un-aliased, free and linear.
\end{example}

Note that an abstract \emph{projection} operator, i.e.,
an operator that actually modifies the set of variables of interest,
is easily specified by composing the operator $\aprojSFL$
with an operator that simply removes, from all the components of $\SFL$
\emph{and} from the set of variables of interest $\VI$,
those variables that have to be projected out.

\section{A Formal Comparison Between $\SFL$ and $\ASub$}
\label{sec:SFL-ASub-comparison}

As we have already observed,
Example~\ref{ex:amguSFL-both-lin-share} shows that
the abstract domain $\SFL$, when equipped with the abstract mgu
operator introduced in Section~\ref{subsec:SFL-operators}, can yield
results that are strictly more precise than all the classical combinations
of set-sharing with freeness and linearity information.
In this section we show that the same example has another interesting,
unexpected consequence, since it can be used to formally prove
that all the classical combinations of set-sharing with freeness
and linearity, including those presented
in~\cite{BagnaraZH00,BruynoogheCM94,HansW92,Langen90th},
are not \emph{uniformly} more precise than the abstract domain
$\ASub$~\cite{Sondergaard86}, which is based on pair-sharing.

To formalize the above observation, we now introduce
the $\ASub$ domain and the corresponding abstract semantics operators
as specified in~\cite{CodishDY91}.
The elements of the abstract domain $\ASub$ have two components:
the first one is a set of variables that are known to be
definitely ground;
the second one encodes both possible pair-sharing and
possible non-linearity into a single relation defined
on the set of variables.
Intuitively,
when $x \neq y$ and $(x,y) \in \VI^2$ occurs in the second component,
then $x$ and $y$ may share a variable;
when $(x,x) \in \VI^2$ occurs in the second component,
then $x$ may be non-linear.
The second component always encodes a symmetric relation; thus,
for notational convenience and without any loss of generality~\cite{King00},
we will represent each pair $(x,y)$ in such a relation as
the sharing group $S = \{x,y\}$,
which will have cardinality 1 or 2 depending on
whether $x = y$ or not, respectively.

\begin{definition} \summary{(The domain $\ASub_{\bot}$.)}
\label{def:ASub}
The abstract domain $\ASub_{\bot}$ is defined as
$\ASub_{\bot} \defeq \{ \botASub \} \union \ASub$, where
\begin{equation*}
  \ASub
    \defeq
      \sset{
        \langle G, R \rangle
          \in
            \wp(\VI) \times \SH
      }{
        G \inters \vars(R) = \emptyset, \\
        \forall S \in R \itc 1 \leq \card S \leq 2
      }.
\end{equation*}
For $i \in \{1, 2\}$, let $\kappa_i = \langle G_i, R_i \rangle \in \ASub$. Then
\[
  \kappa_1 \leqASub \kappa_2
    \quad \defiff \quad
      G_1 \Sseq G_2
        \land
       R_1 \sseq R_2.
\]
The partial order $\mathord{\leqASub}$ is extended on $\ASub_\bot$
by letting $\botASub$ be the bottom element.

Let $u,v \in \VI$ and $\kappa = \langle G, R \rangle \in \ASub$.
Then $u \mayshareoroccurtwice{\kappa} v$ is a shorthand for
the condition $\{u,v\} \in R$,
whereas $u \mayshare{\kappa} v$ is a shorthand for
$u = v \lor \{u,v\} \in R$.
\end{definition}

It is well-known that the domain $\ASub_\bot$ can be obtained by
a further abstraction of any domain such as $\SFL$
that is based on set-sharing and enhanced with
linearity information.
The following definition formalizes this abstraction.

\begin{definition} \summary{($\fund{\abstrASub}{\SFL}{\ASub_\bot}$.)}
\label{def:abstrASub}
Let $\sfl = \langle \sh, f, l \rangle \in \SFL$.
Then
\begin{equation*}
  \abstrASub(\sfl)
    \defeq
      \begin{cases}
        \botASub,
          &\text{if $\sfl = \botSFL$;} \\
        \langle G, R \rangle,
          &\text{otherwise;}
      \end{cases}
\end{equation*}
where
\begin{align*}
  G &\defeq
      \bigl\{\,
        x \in \VI
      \bigm|
        x \notin \vars(\sh)
      \,\bigr\}, \\
  R &\defeq
      \bigl\{\,
        \{ x \} \sseq \VI
      \bigm|
        x \in \vars(\sh) \land x \notin l
      \,\bigr\} \\
    &\quad \quad \union
      \bigl\{\,
        \{ x, y \} \sseq \VI
      \bigm|
        x \neq y
          \land
        \exists S \in \sh \st \{ x, y \} \sseq S
      \,\bigr\}.
\end{align*}
\end{definition}

The definition of abstract unification in~\cite{CodishDY91}
is based on a few auxiliary operators.
The first of these introduces the concept of abstract multiplicity
for a term under a given abstract substitution, therefore modeling
the notion of definite groundness and definite linearity.

\begin{definition} \summary{(Abstract multiplicity.)}
\label{def:chi}
Let $\kappa = \langle G, R \rangle \in \ASub$ and
let $t \in \HTerms$ be a term such that $\vars(t) \sseq \VI$.
We say that
$y \in \vars(t)$ \emph{occurs linearly (in $t$) in $\kappa$}
if and only if
$\fund{\occlin_{\kappa}}{\VI \times \HTerms}{\Bool}$
holds for $(y, t)$, where
\begin{align*}
  \occlin_\kappa(y,t)
    &\defeq
      y \in G
        \lor
          \bigl(
            \occlin(y, t)
              \land
            \forall z \in \vars(t)
              \itc
                \{y,z\} \notin R
          \bigr). \\
\intertext{%
We say that $t$ \emph{has abstract multiplicity $m$ in $\kappa$}
if and only if
$\chi_{\kappa}(t) = m$, where
$\fund{\chi_{\kappa}}{\HTerms}{\{0,1,2\}}$
is defined as follows
}
  \chi_\kappa(t)
    &\defeq
      \begin{cases}
        0, &\text{if $\vars(t) \sseq G$;} \\
        1, &\text{if $\forall y \in \vars(t) \itc \occlin_\kappa(y, t)$;} \\
        2, &\text{otherwise.}
      \end{cases} \\
\intertext{%
For any binding $x \mapsto t$, the function
$\fund{\chi_{\kappa}}{\Bind}{\{0\} \union \{1,2\}^2}$
is defined as follows
}
  \chi_\kappa(x \mapsto t)
    &\defeq
     \begin{cases}
       0,
         &\text{if $\chi_\kappa(x) = 0$ or $\chi_\kappa(t) = 0$;} \\
       \bigl(
         \chi_\kappa(x),
         \chi_\kappa(t)
       \bigr),
         &\text{otherwise;}
     \end{cases}
\end{align*}
\end{definition}
It is worth noting that, modulo a few insignificant differences in
notation, the multiplicity operator $\chi_\kappa$ defined above
corresponds to the abstract multiplicity operator $\chi^\cA$,
which was introduced in~\cite[Definition~3.4]{CodishDY91} and
provided with an executable specification in~\cite[Definition~4.3]{King00}.
Similarly, the next definition corresponds
to~\cite[Definition~4.3]{CodishDY91}.

\begin{definition} \summary{(Sharing caused by an abstract equation.)}
\label{def:soln}
For each $\kappa \in \ASub$
and $(x \mapsto t) \in \Bind$,
where $V_x = \{ x \}$ and $V_t = \vars(t)$ are such that
$V_x \union V_t \sseq \VI$,
the function $\fund{\soln}{\ASub \times \Bind}{\ASub}$
is defined as follows
\begin{align*}
  \soln(\kappa, x \mapsto t)
    &\defeq
      \begin{cases}
        \langle
          V_x \union V_t,
          \emptyset
        \rangle,
          &\text{if $\chi_\kappa(x \mapsto t) = 0$;} \\
        \bigl\langle
          \emptyset,
          \binASub(V_x, V_t)
        \bigr\rangle,
          &\text{if $\chi_\kappa(x \mapsto t) = (1,1)$;} \\
        \bigl\langle
          \emptyset,
          \binASub(V_x, V_x \union V_t)
        \bigr\rangle,
          &\text{if $\chi_\kappa(x \mapsto t) = (1,2)$;} \\
        \bigl\langle
          \emptyset,
          \binASub(V_x \union V_t, V_t)
        \bigr\rangle,
          &\text{if $\chi_\kappa(x \mapsto t) = (2,1)$;} \\
        \bigl\langle
          \emptyset,
          \binASub(V_x \union V_t, V_x \union V_t)
        \bigr\rangle,
          &\text{if $\chi_\kappa(x \mapsto t) = (2,2)$;}
      \end{cases} \\
\intertext{%
where the function $\fund{\binASub}{\wp(\VI)^2}{\SH}$,
for each $V,W \sseq \VI$, is defined as follows
}
  \binASub(V, W)
    &\defeq
      \bigl\{\,
        \{ v, w \} \sseq \VI
      \bigm|
        v \in V, w \in W
      \,\bigr\}.
\end{align*}
\end{definition}

The next definition corresponds to~\cite[Definition~4.5]{CodishDY91}.
\begin{definition} \summary{(Abstract composition.)}
\label{def:abstract-composition}
Let $\kappa, \kappa' \in \ASub$,
where $\kappa = \langle G, R \rangle$
and $\kappa' = \langle G', R' \rangle$.
Then
\(
  \kappa \compose \kappa'
    \defeq
      \langle G'', R'' \rangle
\),
where
\begin{align*}
  G''
    &\defeq
      G \union G', \\
  R''
    &\defeq
      \sset{
        \{u, v\} \in \SH
      }{
        \{ u, v \} \inters G'' = \emptyset, \\
        \bigl(
          u \mayshareoroccurtwice{\kappa} v
        \bigr)
        \lor
        \bigl(
          \exists x, y \st
            u \mayshare{\kappa} x
              \mayshareoroccurtwice{\kappa'}
            y \mayshare{\kappa} v
        \bigr)
      }.
\end{align*}
\end{definition}

We are now ready to define the abstract $\mgu$ operator for the domain
$\ASub_\bot$. This operator can be viewed as a specialization
of~\cite[Definition~4.6]{CodishDY91}
for the case when we have to abstract a single binding. 

\begin{definition} \summary{(Abstract $\mgu$ for $\ASub_{\bot}$.)}
\label{def:amguASub}
Let $\kappa \in \ASub_{\bot}$ and $(x \mapsto t) \in \Bind$,
where $\{x\} \union \vars(t) \sseq \VI$.
Then
\begin{equation*}
  \amguASub(\kappa, x \mapsto t)
    \defeq
      \begin{cases}
        \botASub,
          &\text{if $\kappa = \botASub$;} \\
        \kappa \compose \soln(\kappa, x \mapsto t),
          &\text{otherwise;}
      \end{cases}
\end{equation*}
\end{definition}

By repeating the abstract computation of
Example~\ref{ex:amguSFL-both-lin-share} on the domain $\ASub$,
we provide a formal proof that all the classical approaches
based on set-sharing are not uniformly more precise than
the pair-sharing domain $\ASub$.

\begin{example}
\label{ex:amguASub-both-lin-share}
Consider the substitutions $\sigma,\tau \in \RSubst$
and the abstract element $\sfl \in \SFL$
as introduced in Example~\ref{ex:amguSFL-both-lin-share}.

By Definition~\ref{def:abstrASub}, we obtain
\(
  \kappa
    = \abstrASub( \sfl )
    = \langle \emptyset, R \rangle
\),
where
\begin{align*}
  R &= \{ xx_1, xx_2, xy, xz, yy_1, yy_2, yz \},
\end{align*}

When abstractly evaluating the binding $x \mapsto y$ according to
Definition~\ref{def:amguASub}, we compute the following:
\begin{align*}
  \chi_\kappa(x \mapsto y)
    &= (1,1), \\
  \soln(\kappa, x \mapsto y)
    &= \bigl\langle
         \emptyset,
         \{ xy \}
       \bigr\rangle, \\
  \amguASub(\kappa, x \mapsto y)
    &= \kappa \compose \soln(\kappa, x \mapsto y)
    = \langle \emptyset, R'' \rangle,
\end{align*}
where
\[
  R''
    = R \union
          \{ x, xy_1, xy_2,
             x_1y, x_1y_1, x_1y_2, x_1z,
             x_2y, x_2y_1, x_2y_2, x_2z,
             y, y_1z, y_2z, z \}.
\]
Note that $\{x_1, x_2\} \notin R''$ and $\{y_1, y_2\} \notin R''$,
so that these pairs of variables keep their independence.
In contrast, as observed in Example~\ref{ex:amguSFL-both-lin-share},
the operators in~\cite{BagnaraZH00,BruynoogheCM94,HansW92,Langen90th}
will fail to preserve the independence of these pairs.
\end{example}

We now show that the abstract domain $\SFL$,
when equipped with the operators introduced in
Section~\ref{subsec:SFL-operators},
is uniformly more precise than the domain $\ASub$.
In particular, the following theorem states that
the abstract operator $\amguSFL$ of Definition~\ref{def:amguSFL}
is uniformly more precise than the abstract operator $\amguASub$.
 
\begin{theorem}
\label{thm:amguSFL-uniformly-more-precise-than-amguASub}
Let $\sfl \in \SFL$ and $\kappa \in \ASub_\bot$
be such that $\abstrASub(\sfl) \leqASub \kappa$.
Let also $(x \mapsto t) \in \Bind$,
where $\{x\} \union \vars(t) \sseq \VI$.
Then
\[
  \abstrASub\bigl( \amguSFL(\sfl, x \mapsto t) \bigr)
    \leqASub
      \amguASub(\kappa, x \mapsto t).
\]
\end{theorem}
Similar results can be stated for the other abstract operators,
such as the abstract existential quantification $\aprojSFL$
and the merge-over-all-path operator $\alubSFL$.
It is worth stressing that, when sequences of bindings come into play,
the specification provided in~\cite[Definition~4.7]{CodishDY91}
requires that the \emph{grounding} bindings
(i.e., those bindings such that $\chi_\kappa(x \mapsto t) = 0$)
are evaluated before the non-grounding ones.
Clearly, if we want to lift the result of
Theorem~\ref{thm:amguSFL-uniformly-more-precise-than-amguASub}
so that it also applies to the operator $\aunifySFL$,
the same evaluation strategy has to be adopted when computing
on the domain $\SFL$;
this improvement is well-known~\cite[pp.~66-67]{Langen90th}
and already exploited in most implementations
of sharing analysis~\cite{BagnaraZH00}.

\section{$\PSDFL$: Eliminating Redundancies}
\label{sec:SFL-redundant}

As done in~\cite{BagnaraHZ02TCS,ZaffanellaHB02TPLP}
for the plain set-sharing domain $\SH$,
even when considering the richer domain $\SFL$ it is natural to question
whether it contains redundancies with respect to the computation
of the observable properties.

It is worth stressing that the results presented
in~\cite{BagnaraHZ02TCS} and~\cite{ZaffanellaHB02TPLP}
cannot be simply inherited by the new domain.
The concept of ``redundancy'' depends on both the starting domain
and the given observables:
in the $\SFL$ domain both of these have changed.
First of all,
as can be seen by looking at the definition of $\amguSFL$,
freeness and linearity positively interact in the computation
of sharing information:
\emph{a priori} it is an open issue whether or not
the ``redundant'' sharing groups can play a role
in such an interaction.
Secondly, since freeness and linearity information
can be themselves usefully exploited in a number of applications
of static analysis (e.g., in the optimized implementation
of concrete unification or in occurs-check reduction),
these properties have to be included in the observables.

We will now show that the domain $\SFL$
can be simplified by applying the same notion of redundancy
as identified in \cite{BagnaraHZ02TCS}.
Namely, in the definition of $\SFL$ it is possible
to replace the set-sharing component $\SH$ by $\PSD$
without affecting the precision on
groundness, independence, freeness and linearity.
In order to prove such a claim,
we now formalize the new observable properties.
\begin{definition}\summary{(The observables of $\SFL$.)}
\label{def:SFL-observables}
The (overloaded) \emph{groundness} and \emph{independence} observables
$\rhoCon, \rhoPS \in \uco(\SFL)$ are defined,
for each $\langle \sh, f, l \rangle \in \SFL$, by
\begin{align*}
  \rhoCon\bigl( \langle \sh, f, l \rangle \bigr)
    &\defeq
      \bigl\langle
        \rhoCon(\sh),
        \emptyset,
        \emptyset
      \bigr\rangle, \\
  \rhoPS\bigl( \langle \sh, f, l \rangle \bigr)
    &\defeq
      \bigl\langle
        \rhoPS(\sh),
        \emptyset,
        \emptyset
      \bigr\rangle; \\
\intertext{%
the \emph{freeness} and \emph{linearity} observables
$\rhoF, \rhoL \in \uco(\SFL)$ are defined,
for each $\langle \sh, f, l \rangle \in \SFL$, by
}
  \rhoF\bigl( \langle \sh, f, l \rangle \bigr)
    &\defeq
      \langle \SG, f, \emptyset \rangle, \\
  \rhoL\bigl( \langle \sh, f, l \rangle \bigr)
    &\defeq
      \langle \SG, \emptyset, l \rangle.
\end{align*}
\end{definition}

The overloading of $\rhoPSD$ working on the domain $\SFL$
is the straightforward extension of the corresponding operator on $\SH$:
in particular, the freeness and linearity components are left untouched.
\begin{definition}\summary{(Non-redundant $\SFL$.)}
\label{def:rhoPSD-for-SFL}
For each $\langle \sh, f, l \rangle \in \SFL$,
the operator $\rhoPSD \in \uco(\SFL)$ is defined by
\begin{equation*}
  \rhoPSD\bigl( \langle \sh, f, l \rangle \bigr)
    \defeq
      \bigl\langle \rhoPSD(\sh), f, l \bigr\rangle.
\end{equation*}
This operator induces the lattice
\(
  \PSDFL \defeq \rhoPSD(\SFL)
\).
\end{definition}
As proved in~\cite{ZaffanellaHB02TPLP},
we have that $\rhoPSD \ucoleq (\rhoCon \ucomeet \rhoPS)$;
by the above definitions, it is also clear that
$\rhoPSD \ucoleq (\rhoF \ucomeet \rhoL)$;
thus, $\rhoPSD$ is more precise than
the reduced product
$(\rhoCon \ucomeet \rhoPS \ucomeet \rhoF \ucomeet \rhoL)$.
Informally, this means that the domain $\PSDFL$
is able to \emph{represent} all of our observable properties
without precision losses.

The next theorem shows that $\rhoPSD$ is a congruence
with respect to the $\aunifySFL$, $\alubSFL$ and $\aprojSFL$ operators.
This means that the domain $\PSDFL$ is able
to \emph{propagate} the information on the observables
as precisely as $\SFL$,
therefore providing a completeness result.

\begin{theorem}
\label{thm:PSD-precise-for-SFL}
Let $\sfl_1, \sfl_2 \in \SFL$ be such that
$\rhoPSD(\sfl_1) = \rhoPSD(\sfl_2)$.
Then, for each sequence of bindings $\bs \in \Bind^\ast$,
for each $\sfl' \in \SFL$ and $V \in \wp(\VI)$,
\begin{align*}
     \rhoPSD\bigl(\aunifySFL(\sfl_1, \bs)\bigr)
  &= \rhoPSD\bigl(\aunifySFL(\sfl_2, \bs)\bigr), \\
     \rhoPSD\bigl(\alubSFL(\sfl_1, \sfl')\bigr)
  &= \rhoPSD\bigl(\alubSFL(\sfl_2, \sfl')\bigr), \\
     \rhoPSD\bigl(\aprojSFL(\sfl_1, V)\bigr)
  &= \rhoPSD\bigl(\aprojSFL(\sfl_2, V)\bigr).
\end{align*}
\end{theorem}

Finally, by providing the minimality result,
we show that the domain $\PSDFL$
is indeed the generalized quotient~\cite{CortesiFW98,GiacobazziSR98b}
of $\SFL$ with respect to the reduced product
\(
  (\rhoCon \ucomeet \rhoPS \ucomeet \rhoF \ucomeet \rhoL).
\)

\begin{theorem}
\label{thm:PSD-aunifySFL:minimality}
For each $i \in \{1,2\}$,
let $\sfl_i = \langle \sh_i, f_i, l_i \rangle \in \SFL$
be such that $\rhoPSD(\sfl_1) \neq \rhoPSD(\sfl_2)$.
Then there exist a sequence of bindings $\bs \in \Bind^\ast$
and an observable property $\rho \in \{ \rhoCon, \rhoPS, \rhoF, \rhoL \}$
such that
\[
  \rho\bigl(\aunifySFL(\sfl_1, \bs)\bigr)
    \neq
      \rho\bigl(\aunifySFL(\sfl_2, \bs)\bigr).
\]
\end{theorem}

As far as the implementation is concerned,
the results proved
in~\cite{BagnaraHZ02TCS}
for the domain $\PSD$ can also be applied to $\PSDFL$.
In particular,
in the definition of $\amguSFL$
every occurrence of the star-union operator
can be safely replaced by the self-bin-union operator.
As a consequence, it is possible to provide an implementation
where the time complexity of the $\amguSFL$ operator
is bounded by a polynomial in the number of sharing groups
of the set-sharing component.

The following result provides another optimization
that can be applied when both terms $x$ and $t$ are
definitely linear, but none of them is definitely free
(i.e., when we compute $\sh''$ by the second case
stated in Definition~\ref{def:amguSFL}).

\begin{theorem}
\label{thm:PSD-precise-for-SFL:inner-bin-unions-useless}
Let $\sh \in \SH$ and $(x \mapsto t) \in Bind$,
where $\{x\} \union \vars(t) \sseq \VI$.
Let 
$\sh_{-} \defeq \irel\bigl(\{x\} \union \vars(t), \sh\bigr)$,
$\sh_x \defeq \rel\bigl(\{x\}, \sh\bigr)$,
$\sh_t \defeq \rel\bigl(\vars(t), \sh\bigr)$,
$\sh_{xt} \defeq \sh_x \inters \sh_t$,
$\sh_W \defeq \rel(W, \sh)$, where $W = \vars(t) \setdiff \{x\}$,
and
\[
  \sh^\diamond
    \defeq
      \bin\bigl(
            \sh_x \union \bin(\sh_x, \sh_{xt}^\star),
            \sh_t \union \bin(\sh_t, \sh_{xt}^\star)
          \bigr).
\]
Then it holds
\[
  \rhoPSD\bigl(\cyclicreduce_x^t(\sh_{-} \union \sh^\diamond)\bigr)
    = \begin{cases}
        \rhoPSD\bigl(
                 \sh_{-} \union \bin(\sh_x, \sh_t)
               \bigr),
          &\text{if $x \notin \vars(t)$;} \\
        \rhoPSD\bigl(
                 \sh_{-} \union \bin(\sh_x^2, \sh_W)
               \bigr),
          &\text{otherwise.}
      \end{cases}
\]
\end{theorem}
Therefore, even when terms $x$ and $t$ possibly share
(i.e., when $\sh_{xt} \neq \emptyset$),
by using $\PSDFL$ we can avoid the expensive computation
of at least one of the two inner binary unions
in the expression for $\sh^\diamond$.

\section{Experimental Evaluation}
\label{sec:exp-eval}

Example~\ref{ex:amguSFL-both-lin-share} shows that an analysis
based on the new abstract unification operator can be strictly
more precise than one based on the classical proposal.
However, that example is artificial and leaves open the question
as to whether or not such a phenomenon actually happens during the analysis
of real programs and, if so, how often.
This was the motivation for the experimental evaluation we describe
in this section.
We consider the abstract domain
$\Pos \times \PSDFL$~\cite{BagnaraZH01TPLP},
where the non-redundant version $\PSDFL$ of the domain $\SFL$
is further combined, as described in~\cite[Section~4]{BagnaraZH01TPLP},
with the definite groundness information computed by $\Pos$
and compare the results using the (classical) abstract unification operator
of~\cite[Definition~4]{BagnaraZH01TPLP}
with the (new) operator $\amguSFL$
given in Definition~\ref{def:amguSFL}.
Taking this as a starting point,
we experimentally evaluate eight variants of the analysis arising from
all possible combinations of the following options:
\begin{enumerate}
\item
the analysis can be goal independent or goal dependent;
\item
the set-sharing
component may or may not have widening enabled~\cite{ZaffanellaBH99};
\item
the abstract domain may or may not be upgraded
with structural information using
the $\Pattern(\cdot)$ operator (see~\cite{BagnaraHZ00}
and~\cite[Section~5]{BagnaraZH01TPLP}).
\end{enumerate} 

The experiments have been conducted using the
\china{} analyzer~\cite{Bagnara97th} on a GNU/Linux PC system.
\china{} is a data-flow analyzer for (constraint) logic programs
performing bottom-up analysis and deriving information on both
call-patterns and success-patterns by means of program transformations
and optimized fixpoint computation techniques.
An abstract description is computed for the call- and success-patterns
for each predicate defined in the program.
The benchmark suite,
which is composed of 372 logic programs of various sizes and complexity,
can be considered representative.

The precision results for the goal independent comparisons are summarized
in Table~\ref{tab:oldmodes-vs-newmodes-precision}.
For each benchmark, precision is measured by counting
the number of independent pairs as well as the numbers of
definitely ground, free and linear variables detected.
For each variant of the analysis, these numbers are then compared
by computing the relative precision improvements and expressing them
using percentages.
The benchmark suite is then partitioned into several
precision equivalence classes
and the cardinalities of these classes are shown
in Table~\ref{tab:oldmodes-vs-newmodes-precision}.
For example, when considering a goal independent analysis
without structural information and without widenings,
the value 5 found at the intersection
of the row labeled `$0 < p \leq 2$' with the column labeled `I'
should be read:
``for five benchmarks there has been a (positive) increase in the number
of independent pairs of variables which is less than or equal to two percent.''
Note that we only report on independence and linearity
(in the columns labeled `I' and `L', respectively),
because no differences have been observed for groundness and freeness.
The precision class labeled `unknown' identifies
those benchmarks for which the analyses timed-out
(the time-out threshold was fixed at 600 seconds).
Hence, for goal independent analyses, a precision
improvement affects from 1.6\% to 3\% of the benchmarks,
depending on the considered variant.

When considering the goal dependent analyses,
we obtain a single, small improvement,
so that no comparison tables are included here:
the improvement, affecting linearity information, can be observed
when the abstract domain includes structural information.

With respect to differences in the efficiency,
the introduction of the new abstract unification
operator has no significant effect on the computation time:
small differences (usually improvements) are observed on as many as
6\% of the benchmarks for the goal independent analysis
without structural information and without widenings;
other combinations register even less differences.

We note that it is not surprising that the precision and efficiency
improvements occur very rarely since the abstract unification
operators behave the same except under very specific conditions:
the two terms being unified must not only be definitely linear, but
also possibly non-free and share a variable.

\begin{table*}
\centering
\begin{tabular}{||c||r|r||r|r||r|r||r|r||}
\hhline{~|t:====:t:====:t|}
    \multicolumn{1}{c||}{Goal}
  & \multicolumn{4}{c||}{Without Widening}
  & \multicolumn{4}{c||}{With Widening} \\
\hhline{~|:==:t:==::==:t:==:|}
    \multicolumn{1}{c||}{Independent}
  & \multicolumn{2}{c||}{w/o SI}
  & \multicolumn{2}{c||}{with SI}
  & \multicolumn{2}{c||}{w/o SI}
  & \multicolumn{2}{c||}{with SI} \\
\hhline{|t:=::==::==::==::==:|}
 Prec. class
 & \multicolumn{1}{c|}{I}
 & \multicolumn{1}{c||}{L}
 & \multicolumn{1}{c|}{I}
 & \multicolumn{1}{c||}{L}
 & \multicolumn{1}{c|}{I}
 & \multicolumn{1}{c||}{L}
 & \multicolumn{1}{c|}{I}
 & \multicolumn{1}{c||}{L} \\
\hhline{|:=::==::==::==::==:|}
$\phantom{-1}5 < p \leq 10\phantom{-}$ &
  --- &   2 & --- &   2 & --- &   2 & --- &   2 \\
\hhline{||-||--||--||--||--||}
$\phantom{-1}2 < p \leq 5\phantom{-0}$ &
  --- & --- & --- & --- & --- & --- & --- &   1 \\
\hhline{||-||--||--||--||--||}
$\phantom{-1}0 < p \leq 2\phantom{-0}$ &
    5 &   5 &   9 &   6 &   6 &   6 &  12 &   8 \\
\hhline{||-||--||--||--||--||}
same precision & 
  357 & 355 & 337 & 338 & 366 & 364 & 360 & 361 \\
\hhline{||-||--||--||--||--||}
unknown &
   10 &  10 &  26 &  26 & --- & --- & --- & --- \\
\hhline{|b:=:b:==:b:==:b:==:b:==:b|}
\end{tabular}

\caption{Classical $\Pos \times \PSDFL$ versus enhanced one: precision.}
\label{tab:oldmodes-vs-newmodes-precision}
\end{table*}

\section{Related Work}
\label{sec:related}

Sharing information has been shown to be important for
finite-tree analysis~\cite{BagnaraGHZ01,BagnaraZGH01}.
This aims at identifying those program variables that,
at a particular program point,
cannot be bound to an infinite rational tree
(in other words, they are necessarily bound to acyclic terms).
This novel analysis is irrelevant for those logic languages
computing over a domain of finite trees,
while having several applications for those
(constraint) logic languages that are explicitly
designed to compute over a domain including rational trees,
such as Prolog~II and its successors \cite{Colmerauer82,Colmerauer90},
SICStus Prolog \cite{SICStusManual},
and Oz \cite{SmolkaT94}.
The analysis specified in~\cite{BagnaraGHZ01} is based on
a parametric abstract domain $H \times P$,
where the $H$ component (the Herbrand component)
is a set of variables that are known to be bound to finite terms,
while the parametric component $P$ can be any domain
capturing aliasing, groundness, freeness and linearity information
that is useful to compute finite-tree information.
An obvious choice for such a parameter is the domain
combination $\SFL$.
It is worth noting that, in~\cite{BagnaraGHZ01}, the correctness
of the finite-tree analysis is proved by \emph{assuming}
the correctness of the underlying analysis on the parameter $P$.
Thus, thanks to the results shown in this paper,
the proof for the domain $H \times \SFL$ can now be considered complete.

Codish \etal{}~\cite{CodishLB00} describe
an algebraic approach to the sharing analysis of
logic programs that is based on \emph{set logic programs}.
A set logic program is a logic program
in which the terms are sets of variables and
standard unification is replaced by a suitable unification for sets,
called \emph{ACI1-unification}
(unification in the presence of an associative,
commutative, and idempotent equality theory with a unit element).
The authors show that the domain of \emph{set-substitutions},
with a few modifications, can be used as an abstract domain
for sharing analysis. They also provide an isomorphism between
this domain and the set-sharing domain $\SH$ of Jacobs and Langen.
The approach using set logic programs is also generalized
to include linearity information, by suitably annotating
the set-substitutions, and the authors formally state
the optimality of the corresponding abstract unification operator
$\textit{lin-mgu}_{\textit{ACI1}}$
(Lemma~A.10 in the Appendix of~\cite{CodishLB00}).
However, this operator is very similar to the classical combinations
of set-sharing with linearity~\cite{BruynoogheCM94,HansW92,Langen90th}:
in particular, the precision improvements arising from
this enhancement are only exploited when the two terms being unified
are definitely independent.
As we have seen in this paper,
such a choice results in a sub-optimal abstract unification operator,
so that the optimality result cannot hold.
By looking at the proof of Lemma~A.10 in~\cite{CodishLB00},
it can be seen that the case when the two terms possibly share
a variable is dealt with by referring to an example:\footnote{%
The proof refers to Example 8, which however has nothing to do
with the possibility that the two terms share;
we believe that Example 2 was intended.}
this one is supposed to show that
all the possible sharing groups can be generated.
However, even our improved operator correctly characterizes
the given example, so that the proof is wrong.
It should be stressed that the $\amguSFL$ operator presented
in this paper, though remarkably precise,
is not meant to subsume all of the proposals
for an improved sharing analysis
that appeared in the recent literature
(for a thorough experimental evaluation of many of these proposals,
the reader is referred to~\cite{BagnaraZH00,Zaffanella01th}).
In particular, it is not difficult to show that our operator
is not the optimal approximation of concrete unification.

In a very recent paper~\cite{HoweK03},
J.~Howe and A.~King consider the domain $\SFL$
and propose three optimizations to improve
both the precision and the efficiency of the (classical)
abstract unification operator.
The first optimization is based on the same observation
we have made in this paper, namely that the independence check
between the two terms being unified is not necessary for ensuring
the correctness of the analysis.
However, the proposed enhancement does not fully exploit this observation,
so that the resulting operator is strictly less precise than
our $\amguSFL$ operator (even when the operator $\cyclicreduce_x^t$
does not come into play).
In fact, the first optimization of~\cite{HoweK03}
is not uniformly more precise than the classical proposals.
The following example illustrates this point.

\begin{example}
\label{ex:amguHW-less-precise-than-classical-amgu}
Let $\VI = \{x, y, z_1, z_2, z_3\}$, $(x \mapsto y) \in \Bind$ and
\(
  \sfl
    \defeq \langle \sh, \emptyset, \VI \rangle
\),
where
\(
  \sh = \{ xz_1, xz_2, xz_3, yz_1, yz_2, yz_3 \}
\).

Since $x$ and $y$ are linear and independent,
$\amguSFL$ as well as
all the classical abstract unification operators will compute
\(
  \sfl_1 = \bigl\langle \sh_1, \emptyset, \{x, y\} \bigr\rangle
\),
where
\[
  \sh_1 \defeq \bin(\sh_x, \sh_y) 
        = \{ xyz_1, xyz_1z_2, xyz_1z_3, xyz_2, xyz_2z_3, xyz_3 \}.
\]
In contrast,
a computation based on~\textup{\cite[Definition~3.2]{HoweK03}},
results in the less precise abstract element
\(
  \sfl_2 = \bigl\langle \sh_2, \emptyset, \{x, y\} \bigr\rangle
\),
where
\[
  \sh_2 \defeq \bin(\sh_x^\star, \sh_y) \inters \bin(\sh_x, \sh_y^\star)
        = \sh_1 \union \{ xyz_1z_2z_3 \}.
\]
\end{example}

The second optimization shown in~\cite{HoweK03} is based
on the enhanced combination of set-sharing and freeness information,
which was originally proposed in~\cite{File94}.
In particular, the authors propose a slightly different precision
enhancement, less powerful as far as precision is concerned,
which however seems to be amenable for an efficient implementation.
The third optimization in~\cite{HoweK03} exploits the combination
of the domain $\SFL$ with the groundness domain $\Pos$.

\section{Conclusion}
\label{sec:conclusion}

In this paper
we have introduced the abstract domain $\SFL$, combining
the set-sharing domain $\SH$ with freeness and linearity information.
While the carrier of $\SFL$ can be considered standard,
we have provided the specification of a new abstract unification operator,
showing examples where this operator achieves more precision
than the classical proposals.
The main contributions of this paper are the following:
\begin{itemize}
\item
we have defined a precise abstraction function,
mapping arbitrary substitutions in rational solved form
into their \emph{most precise} approximation on $\SFL$;
\item
using this abstraction function,
we have provided the mandatory proof of \emph{correctness}
for the new abstract unification operator,
\emph{for both finite-tree and rational-tree languages};
\item
we have formally shown that the domain $\SFL$ is
\emph{uniformly} more precise than the domain $\ASub$;
we have also provided an example showing that all the classical
approaches to the combinations of set-sharing with freeness and
linearity fail to satisfy this property;
\item
we have shown that, in the definition of $\SFL$,
we can replace the set-sharing domain $\SH$
by its non-redundant version $\PSD$.
As a consequence, it is possible to implement an algorithm
for abstract unification running in \emph{polynomial time}
and still obtain the same precision on all the considered observables,
that is groundness, independence, freeness and linearity.
\end{itemize}

\section*{Acknowledgment}

We recognize the hard work required to review technical papers
such as this one and would like to express our real gratitude
to the Journal referees for their critical reading and constructive
suggestions for preparing this improved version.

\hyphenation{ Ba-gna-ra Bie-li-ko-va Bruy-noo-ghe Common-Loops DeMich-iel
  Dober-kat Er-vier Fa-la-schi Fell-eisen Gam-ma Gem-Stone Glan-ville Gold-in
  Goos-sens Graph-Trace Grim-shaw Her-men-e-gil-do Hoeks-ma Hor-o-witz Kam-i-ko
  Kenn-e-dy Kess-ler Lisp-edit Lu-ba-chev-sky Nich-o-las Obern-dorf Ohsen-doth
  Par-log Para-sight Pega-Sys Pren-tice Pu-ru-sho-tha-man Ra-guid-eau Rich-ard
  Roe-ver Ros-en-krantz Ru-dolph SIG-OA SIG-PLAN SIG-SOFT SMALL-TALK Schee-vel
  Schlotz-hauer Schwartz-bach Sieg-fried Small-talk Spring-er Stroh-meier
  Thing-Lab Zhong-xiu Zac-ca-gni-ni Zaf-fa-nel-la Zo-lo }

\end{document}